\documentclass[11pt,a4paper]{article}
\usepackage{jcappub}
\usepackage{amsmath}
\usepackage{amsfonts}
\usepackage{amssymb}
\usepackage[utf8]{inputenc}
\usepackage{graphicx}

\title{Covariant Effective Action for Scalar-Tensor Theories of Gravity}

\author[a]{Sandeep Aashish,}
\author[b]{Sukanta Panda,}
\author[b]{Abbas Altafhussain Tinwala,}
\author[b]{and Archit Vidyarthi}

\affiliation[a]{Department of Physics, School of Advanced Sciences, Kalasalingam Academy of Research and Education, Krishnankoil - 626126, India}
\affiliation[b]{Department of Physics, Indian Institute of Science Education and Research Bhopal - 462066, India}

\emailAdd{sandeepaashish@klu.ac.in}
\emailAdd{sukanta@iiserb.ac.in}
\emailAdd{abbas18@iiserb.ac.in}
\emailAdd{archit17@iiserb.ac.in}

\abstract{We develop the calculation of the divergent part of one-loop covariant effective action for scalar fields minimally and non-minimally coupled to gravity using the generalized Schwinger-DeWitt technique. We derive the field-space metric using Vilkovisky's prescription and obtain new terms in the one-loop corrections which are absent in past literature with trivial choices of field-space metric. We further calculate the covariant versions of past results, obtained using the standard approach, by applying Barvinsky and Vilkovisky's extension to the ordinary Schwinger-DeWitt approach. For completeness, we study the one-loop gravitational corrections for a dilaton-extended two-field Starobinsky model and compare with known results.}

\arxivnumber{2104.12713}

\begin{document}
\maketitle
\flushbottom

\section{Introduction}
    Much of the recent work in the field of theoretical physics has been aimed towards uniting gravity with the rest of known matter under the shared umbrella of a unified theory. Despite all efforts to reconcile them, however, the solution to this long-standing problem, owing to the non-renormalizability of gravity, has so far proved elusive \cite{wood}. It is a viewpoint shared by several in the community that perhaps the best way to go about the problem would be to work with a quantum effective field theory of gravity \cite{donoghue}, which allows us to separate the high energy quantum phenomena from the low energy general relativity results that have been verified extensively over the past century.
    
	One area that would benefit the most from a unified theory is cosmology. With the success of the inflation model in explaining both CMB observations and large-scale structure formation, it has become increasingly evident that quantum fluctuations at the beginning of inflation would have played a huge role in the formation of the universe as we see it today \cite{quantcorr,wilczek,wooda}. Leading order quantum corrections to classical gravity have been able to explain the aforementioned problems in inflationary scenarios to a great extent \cite{klemm,gog,gog2,herr,markk,iaon,ruf,inst,higg}, and so effective theories of gravity are expected to forego the renormalizability constraint, and instead focus on the corrections that are `relevant' at the energy scale in question. 
	
	Effective action is known to be the generator of one-particle irreducible (1PI) diagrams, which reduces to the classical action in the background field limit, and contains all information about the quantum fluctuations in the form of a perturbative expansion where each order of correction corresponds to the number of loops \cite{diag,vilk}. The effective action is, consequently, a functional of the background field \cite{falken,nonlineargauges}, and it was with the aim of developing a field-reparametrization invariant and gauge-condition independent formalism, that Vilkovisky-DeWitt's (VDW) covariant effective action approach was introduced \cite{unique,lavrov, odin,effbook}, with a generalization to composite fields in Ref. \cite{odin2}. Some works that best demonstrate the strengths of this technique in various settings are Refs. \cite{evendim,kaluza,feynman,ym,conform}. From this point forward, we shall label the standard effective action as `non-covariant' with respect to reparametrizations in the field space.
	
	The primary difference between the non-covariant effective action and the VDW effective action is the off-shell covariance of results. When calculating on-mass-shell, the results obtained by the background field approach of non-covariant effective action match exactly with those obtained by VDW effective action, since gauge condition independence is akin to reparametrization invariance, which is vital at the classical level according to Ref. \cite{borcher}. However, the covariant results are more suited for off-shell calculations, where virtual phenomena come into the picture. For a more detailed description of the difference, please refer to Ref. \cite{unique}. Some recent works that highlight the important properties of VDW effective action are Refs. \cite{giacchini1,giacchini2}.
	
	The VDW approach reformulates the problem in terms of the space of fields, where reparametrization invariance becomes equivalent to a general coordinate transformation in the configuration-space manifold, which in turn ensures covariance of the approach through Vilkovisky-DeWitt connections. The main issue encountered in this formalism is in the choice of configuration-space metric \cite{odin3,odin4}. While Vilkovisky advocated obtaining the operator from the highest derivative term in the action, several past works like Refs. \cite{mackay,gali,ashish} have instead explicitly chosen diagonal metric forms that seem easier to deal with. The calculations consequently yield inaccurate results with missing terms attributed to the off-diagonal components of the metric. The paper \cite{karam1} lists an explicit formula following Vilkovisky's prescription for finding the field-space metric, and the authors go a step forward and exclude dependence on total derivative terms present in the Lagrangian in a subsequent work \cite{karam2}.
	
	At present, the most commonly used method for the evaluation of VDW effective action involves perturbative expansion in orders of background fields, followed by calculating the loop integrals, as outlined in Ref. \cite{mackay}. However, there exists an alternative called the heat kernel technique, where one can obtain closed-form results for the loop corrections to the effective action. It initially found application in quantum field theory from the works of Fock \cite{fock} and Schwinger \cite{schw}, who noticed that the Green functions can be represented as integrals over an auxiliary `proper time' variable \cite{schwartz}. In a subsequent work, DeWitt extended the technique to general non-trivial backgrounds, thereby completing a powerful tool for computing one–loop divergences in quantum gravity in a manifestly covariant approach \cite{dewittdynamical,dewitt2}.
	
	Since the Green function for a theory corresponds to the differential operator obtained from the action, it is straightforward to see that the form of the operator plays a large part in the heat kernel expansions that we require. Although originally built around `minimal' operators which simply involve a D'Alembertian operator and a potential term, the formalism has since been extended by Barvinsky and Vilkovisky in Ref. \cite{nonminimal} to include a far wider class of `non-minimal' operators, some of which have been covered as part of our calculations in this paper. Such cases have also been considered using the traditional (non-covariant) methods in the past, for example in Refs. \cite{karmazin,takata}.
	
	So far, the attempts at evaluating the one-loop covariant effective action have not been entirely successful. In Ref. \cite{stein}, the calculations were performed using a non-covariant method which was evident due to lack of Vilkovisky-DeWitt connections, while in Refs. \cite{mackay,gali,ashish}, the explicit choice of a trivial field-space metric led to inaccurate results. In this paper, we perform these calculations using the extended Schwinger-DeWitt approach and compare them with existing literature. As an added check, we shall also extract the on-shell effective action from the models that we use and verify the results using Ref. \cite{steinJE}, reaffirming that irrespective of whether we employ a covariant or a non-covariant method, the on-shell results must remain unchanged.
	
	The general layout of the paper is as follows: Section \hyperref[sec2]{2} is dedicated to the Effective Action approach, which briefly introduces the concept of non-covariant effective action and later extends it to the covariant VDW form, with special emphasis on the case of gauge theories as they play an important role in this work; Section \hyperref[heatk]{3} explains in brief the Heat Kernel method, with first an introduction to the technique for `minimal' operators, followed by a subsequent extension to `non-minimal' operators as proposed by Barvinsky and Vilkovisky \cite{nonminimal}; we, then, begin the calculations in Section \hyperref[model]{4} starting with a single real scalar field model coupled both minimally and non-minimally with gravity, and an explicit calculation for a Starobinsky model with an additional scalar field; followed by a renormalizability analysis of the aforementioned models in Section \hyperref[renorm]{5}. We perform various verifications in Section \ref{verify}, wherein we compare non-covariant divergences in one-loop effective action with the existing literature and confirm that both covariant and non-covariant divergences in one-loop effective action lead to the same on-shell result. Finally, we make some concluding remarks in Section \hyperref[result]{7}.

    \section{Effective Action}\label{sec2}
    Schwinger action principle states that
    \begin{equation}
        \delta \left<\text{out}|\text{in}\right>=\frac{i}{\hbar}\left<\text{out}|\delta S|\text{in}\right>
    \end{equation}
    where $\delta S$ includes variations in the action functional w.r.t. variations in fields or any parameters on which the action depends. Here, we can define a functional $W$ such that,
    \begin{equation}
        \left<\text{out}|\text{in}\right>=\exp\left(\frac{i}{\hbar}W\right)
    \end{equation}
    which is identified as the energy functional. Assuming a background field approach, wherein we can break the field into a classical background and its quantum fluctuations $(\phi=\Bar{\phi}+\delta\phi)$, the functional dependence of $W$ on the variation in fields appears to be integrated out. If there is an external source $J$ present in the total action, this definition implies that the energy functional depends only on $J$. Performing a Legendre transform, we get
    \begin{equation}
        \Gamma[\Bar{\phi}]=W[J]-\int dv_x J(x)\Bar\phi(x)
    \end{equation}
    where the functional $\Gamma[\Bar{\phi}]$ is dependent only on the background value of the field $\Bar{\phi}$, and is called the effective action. In the absence of an external source, the energy functional $W$ can be identified as the effective action. For a general (bosonic-field) theory, the one-loop correction to the effective action is given by,
    \begin{equation}\label{eff}
        \Gamma^{(1)} = \frac{i\hbar}{2} \ln\det F(\nabla)
    \end{equation}
    where $F(\nabla)$ represents the differential operator obtained from the action. Contrary to the partition function, which acts as the generator for all possible Feynman diagrams, the energy functional and the effective action functional act as generators for only the connected and 1PI diagrams, respectively \cite{diag,effbook}.
    
    \subsection{VDW Effective Action}
    In Ref. \cite{borcher}, Borchers states that the $\hat{S}$-matrix remains unchanged under field reparametrizations, since they preserve the asymptotic values of fields in their in- and out- regions. And since the effective action reduces to the classical action functional at zeroth order, we expect that imposing the condition of reparametrization invariance would cause no loss of generality.\\
    In path integral form, effective action for a general theory of $N$-bosonic fields can be expressed as the recursion relation $(i=1,..,N)$,
    \begin{equation}\label{vdw}
	    	\exp{\frac{i}{\hbar} \Gamma [\bar{\phi}]} = \int [\mathcal{D} \phi] \exp{\frac{i}{\hbar}\left\{S[\phi] -(\phi^i- \bar{\phi}^i)\frac{\delta\Gamma[\bar{\phi}]}{\delta \bar{\phi}^i}\right\}}
    \end{equation}
    in DeWitt notation. This notation can be easily visualized with the following example: Consider a theory involving fields $\Phi^i=\{\phi^A(x),\phi^B(x),.., A^I_{\mu}(x), A^J_{\mu}(x),.., h_{\mu\nu}(x)\}$. In this representation, the index $i$ holds  information about the spacetime coordinates ($x$, $y$,..), spacetime indices (represented by Greek indices like $\mu$, $\nu$,..), and field space indices (represented by capitalised Latin characters $I$, $J$,..). Summing over indices in this notation additionally implies integrals over the `summed-over' spacetime points as well:
    \begin{align*}
        \Phi^i\Phi_i=&\int dv_x\, \phi^A(x)\phi_A(x) +\int dv_x\, \phi^B(x)\phi_B(x)+...\\ &+ \int dv_x \,A^{I\mu}(x)A_{I\mu}(x) + \int dv_x \,A^{J\mu}(x)A_{J\mu}(x)+... \;+ \int dv_x\, h^{\mu\nu}(x)h_{\mu\nu}(x) 
    \end{align*}
    
    Now, the relation (\hyperref[vdw]{2.5}), due to the factor of $(\phi^i-\bar{\phi}^i)$, is not covariant under a general coordinate transformation in the space of fields. So, to bring it to a more desirable form, Vilkovisky \cite{unique} formulated the unique effective action, which was modified further by DeWitt \cite{dewitt}, resulting in,
    \begin{equation}
         \exp{\frac{i}{\hbar}\Gamma[\bar{\phi};\phi_*]} = \int \prod_id\phi^i |g[\phi]|^\frac{1}{2} \left|\Delta[\phi_*;\phi]\right|
         \exp{\frac{i}{\hbar}\left\{S[\phi] + \frac{\delta\Gamma[\bar{\phi};\phi_*]}{\delta\sigma^i[\phi_*;\bar{\phi}]}\left(\sigma^i[\phi_*;\bar{\phi}] - \sigma^i[\phi_*;\phi]\right)\right\}}
    \end{equation}
    Here, the factor $(\phi^i-\bar{\phi}^i)$ was replaced with
    \begin{equation*}
        \sigma^i[\phi_*;\phi] = g^{ij}[\phi_*]\frac{\delta}{\delta \phi_*^j}\sigma[\phi_*;\phi]
    \end{equation*}
    using Synge's world function $\sigma[\phi_*;\phi]$. $\Delta[\phi_*;\phi]$ is the Van Vleck-Morette determinant, where the choice of $\phi_*$ is arbitrary, and $g[\phi]$ represents the metric of the $N$-dimensional configuration space at point $\phi$, which according to Ref. \cite{metric}, can be read off of the highest even order derivative term in the Lagrangian. DeWitt assigned $\phi_* = \bar{\phi}$, resulting in what is called the DeWitt effective action
    \begin{equation}
	\exp{\frac{i}{\hbar}\Gamma_D[\bar{\phi}]} = \int \prod_id\phi^i |g[\phi]|^\frac{1}{2} \left|\Delta[\bar{\phi};\phi]\right| \exp{\frac{i}{\hbar} \left\{S[\phi] -
	\sigma^i[\bar{\phi};\phi]C^{-1j}_i \frac{\delta\Gamma_D[\bar{\phi}]}{\delta\bar{\phi}^j}\right\}}
	\end{equation}
	where $C^i_j=\left<\sigma^i_{;j}[\bar{\phi};\phi]\right>$, covariant derivatives are taken w.r.t. the field-space connection, and undashed index in the derivative represents that the derivative is w.r.t. the first argument. Also, $\left<A\right>$ represents the statistical average of some functional $A$ w.r.t. the action. Expanding perturbatively, we arrive at the one-loop correction to the effective action,
	\begin{equation}
	    \Gamma^{(1)}[\bar{\phi};\phi_*]=\frac{i\hbar}{2}\ln\det S^{;i}_j
	\end{equation}
    which appears to be the same as the result obtained in the non-covariant case (\hyperref[eff]{2.4}), except for a covariant derivative instead of an partial derivative w.r.t. fields, and a raised index $i$. Both these differences are attributed to a non-trivial field-space metric used in VDW approach, further stressing the importance of following Vilkovisky's prescription for the same \cite{odin3}.
    
    \subsection{Case of Gauge Theories}
    For gauge theories, the action functional remains invariant under transformations belonging to the gauge group. In order to make meaningful calculations, we need to fix the gauge and pick one value of the field from each equivalence class under the gauge group. The field-space under consideration is thus effectively modified, which means corresponding changes are required in the formula for effective action as well. The gauge transformations may be written as
    \begin{equation}\label{gtran}
        \delta \phi^i = K^i_\alpha \delta \epsilon^\alpha
    \end{equation}
    where $\delta\epsilon^\alpha$ are parameters that characterise the transformations and $\mathbf{K}_\alpha=K^i_\alpha\frac{\delta}{\delta\phi^i}$  are vectors in field-space that act as generators of the transformations, such that they form a Lie algebra
    \begin{equation}
        \left[\mathbf{K}_\alpha[\phi],\mathbf{K}_\beta[\phi]\right]=-f_{\alpha\beta}^\gamma \mathbf{K}_\gamma[\phi]
    \end{equation}
    where $f_{\alpha\beta}^\gamma$ are the structure constants for the algebra. Following this line of reasoning, the recursion formula (\hyperref[vdw]{2.5}) is modified to
    \begin{equation}\label{gaugevdw}
        \exp{\frac{i}{\hbar}{\Gamma[v;\phi_*]}} = \int \prod_id\phi^i(\det{g_{ij}})^{1/2} (\det{Q^\alpha_\beta})\tilde{\delta}[\chi^\alpha;0]\exp{\frac{i}{\hbar}\left\{S[\phi]+(v^i-\sigma^i)\frac{\delta}{\delta v^i}\Gamma[v;\phi_*]\right\}}
    \end{equation}
    Here, $v^i = \sigma^i[\phi_*;\bar{\phi}]$, and $\sigma^i = \sigma^i[\phi_*;\phi]$ are chosen to ease typography. $\chi^\alpha=0$ here represent the gauge conditions, and $\tilde{\delta}$ is not the traditional Dirac delta function and is only used to represent the application of the gauge conditions in the integral. Also, $Q^\alpha_\beta =\chi^\alpha_{,i}[\phi]K^i_\beta[\phi]$, and the factor of $\det{Q^\alpha_\beta}\neq 0$ is used to ensure that only one member is picked from each equivalence class.
    
    It was shown in Ref. \cite{effbook} that using Landau-DeWitt gauge conditions: $K^i_\alpha[\phi_*]g_{ij}[\phi_*](\phi^j-\phi^j_*)=0$ for a theory with a field-independent configuration space metric (Yang-Mills theory, for example), we can write $\sigma^i[\phi_*;\phi]=-\eta^i[\phi_*;\phi]=\phi_*^i-\phi^i$, i.e. the configuration space behaves as if it was flat. This changes (\hyperref[gaugevdw]{2.11}) to
    \begin{equation}\label{gaugevdw1}
        \exp{\frac{i}{\hbar}{\Gamma[\bar{\phi};\phi_*]}} = \int \prod_id\phi^i(\det{g_{ij}})^{1/2} (\det{Q^\alpha_\beta})\tilde{\delta}[\chi^\alpha;0]\exp{\frac{i}{\hbar}\left\{S[\phi]+(\phi^i-\bar{\phi}^i)\frac{\delta}{\delta \bar{\phi}^i}\Gamma[\bar{\phi};\phi_*]\right\}}
    \end{equation}
    Using perturbation theory, the one-loop correction to the effective action turns out to be
    \begin{equation}\label{correctcorr}
        \Gamma^{(1)}=-i\hbar\ln\det\bar{Q}_{\alpha\beta}+\frac{i\hbar}{2}\ln\det\bar{S}^{LD}_{,ij}+\frac{i\hbar}{2}\ln\det(-\bar{K}_{\alpha i}\Delta^{ij}\bar{K}_{\beta j})
    \end{equation}
    where overbars mean evaluation at the background value $\bar{\phi}^i$, the superscript $LD$ represents that the action has been gauge fixed, and $\Delta^{ij}$ is the Green function corresponding to $\bar{S}^{LD}_{,ij}$. The last term is a relic from gauge-fixing the integral measure, and makes no non-trivial contribution to the divergent part of the one-loop effective action \cite{effbook}. This result exactly matches what we would have arrived at using the background field method and adopting the Feynman gauge. The configuration-space metric for theories involving gravity is dependent on the spacetime metric field, meaning we cannot reduce (\hyperref[gaugevdw]{2.11}) to (\hyperref[gaugevdw1]{2.12}) in such cases. We can, however, Taylor expand $\sigma^i$ around the flat case $\eta^i=\phi^i-\phi_*^i$ as:
    \begin{equation}\label{synge}
        \sigma^i=-\eta^i-\frac{1}{2}\tilde{\Gamma}^i_{jk}\eta^j\eta^k - \mathcal{O}(\eta^3)
    \end{equation}
    where $\tilde{\Gamma}^i_{jk}$ are the configuration-space Christoffel connections. For one-loop calculations that we intend to perform in this paper, we expect to deal with terms only up to $\mathcal{O}(\eta^2)$ \cite{fradkin} and so, we can still arrive at the result in (\hyperref[correctcorr]{2.13}), albeit with $\bar{S}^{LD}_{;ij}$ involving a covariant derivative instead of a partial derivative.
    
    Since this technique is applicable to theories containing multiple fields, as evident from the form of the field space metric, we shall use $\;\hat{ }\;$ to denote an operator in field space. For example: $\delta^A_{\;B}=\hat 1, P^A_{\;B}=\hat P$.
    
    \section{Heat Kernel Method}\label{heatk}
	\subsection{Schwinger-DeWitt Technique}
	Schwinger \cite{schw} intended to relate Green's function $\hat\Delta$, corresponding to the operator $\hat F(\nabla)$ in (\hyperref[eff]{2.4}), with heat kernel $\hat K(s|x,x')$ which are the general solutions to the heat equation:
	\begin{equation}\label{a3}
		i\frac{\partial}{\partial s}\hat K(s|x,x') = \hat F(\nabla) \hat K(s|x,x')
	\end{equation}
	where $s$ is what Schwinger referred to as the proper time. The relation is given by:
	\begin{equation}\label{a4}
	    \hat G(x,x')=-\hat F^{-1}(\nabla)(x,x')=\int_0^\infty ds\hat K(s|x,x')
	\end{equation}
    This allows us to write,
	\begin{equation}\label{a5}
		W = -\frac{i\hbar}{2}\int dv_x \int_0^{\infty}\frac{ds}{s}\; \text{Tr} \hat K(s|x,x')
	\end{equation}
	In the flat $n$-dimensional spacetime limit, we can expand the heat kernel asymptotically for small $s$ as,
	\begin{equation}\label{a6}
		\hat K(s|x,x') = i(4\pi is)^{-n/2}\sum_{k=0}^{\infty} (is)^k \hat a_k(x,x')
	\end{equation}
	where $\hat a_k(x,x')$ are the HAMIDEW coefficients \cite{dewittdynamical,gibbons,gilkey} and have been calculated previously for a differential operator of the form:
	\begin{equation}\label{min}
		\hat F(\nabla)= g^{\mu\nu}(x)\nabla_\mu\nabla_\nu \;\hat 1+ \hat P(x)
	\end{equation}
	which is the standard `minimal' form of a second order derivative operator, where $\hat{P}(x)$ is some potential term devoid of in the theory. Substituting (\hyperref[a6]{3.4}) in (\hyperref[a5]{3.3}), we observe that the effective action is clearly divergent in the $s$ integral in both limits of the integration $s\rightarrow 0$ (ultraviolet divergent) and $s\rightarrow\infty$ (infrared divergent). Since the asymptotic expansion of the heat kernel (\hyperref[a6]{3.4}) is only possible in the $s\rightarrow 0$ limit, studying the $s\rightarrow\infty$ limit would require a complete reformulation of the approach, known as the late-time asymptotics, and is important in order to understand the infrared properties of the theories under consideration \cite{mukh,nest}. This portion of the expansion is expected to be useful in dealing with the cosmological constant problem \cite{cosmoc}. However, little information is available about this limit in the background field formalism, so we leave this portion as a future exercise. 
	
	The ultraviolet divergences present in the $s\rightarrow 0$ limit shall be the focus of our present work and can be dealt with by using a small, constant value $s_o$ as a cut-off for the integration in the upper limit. Then, the UV divergent part of the effective action becomes,
	\begin{equation}\label{a8}
	    W_{div} = \frac{i\hbar}{2}(4\pi)^{-n/2} \int dv_x \sum_{k=0}^{\infty}\;\text{tr}\, \hat a_k(x,x)
	    \,\left\{\int_0^{s_o}ds (is)^{k-1-n/2}\right\}_{div}
	\end{equation}
	The divergences in this limit are clearly quadratic $(k<\frac{n}{2})$ and logarithmic $(k=\frac{n}{2})$. For $n=4$, these are sourced in the first three coefficients ($k=0, 1, 2$) of the heat kernel expansion (listed in Appendix \hyperref[coeff]{A}). The rest of the summation returns a finite value in the given limits of integration. Further, using dimensional regularization, we find that only the third term in the series contributes to the divergent part of the effective action, meaning only the logarithmic divergences remain.
	
    The spacetime integral in (\hyperref[a8]{3.6}) is usually dealt with by using effective potential rather than the effective action eliminating the need for the spacetime integral. Moving forward, we shall use $\hbar=1$ in all expressions.
	
	\subsection{Non-Minimal Operators}
	Consider a general differential operator,
	\begin{equation}\label{a10}
		\hat F(\nabla) = \hat D(\nabla) + \hat \Pi(\nabla)
	\end{equation}
	where $\hat{ \;}$ represents an operator in the space of fields. Here, $\hat D(\nabla)$ is the $2k^{th}$ order differential operator for some positive integer $k$, while $\hat \Pi(\nabla)$ is a differential operator of order $2k-1$. 
    
    Note that for $k=1$, $\hat F(\nabla)$ is a general second order differential operator but does not necessarily coincide with the standard minimal operator in (\ref{min}). However, a class of such second order operators can be reduced to the minimal form when $\hat D(\nabla)=g^{\mu\nu}(x)\nabla_\mu\nabla_\nu \;\hat 1$ \cite{nonminimal}. 
    The operators which cannot be reduced to the minimal form are called \textit{non-minimal} operators. It is important to make this distinction between minimal and non-minimal operators, because the HAMIDEW coefficients are known for the operator in (\ref{min}) which is infact a minimal operator. Hence, for a theory with minimal operators, the HAMIDEW coefficients are invariably known. For the general differential operator $\hat F(\nabla)$ of order $k$ in (\ref{a10}), the so called \textit{condition of minimality} is given by:
	\begin{equation}\label{a9}
	    \hat D(\nabla)=\hat C(\phi) \;.\;(g^{\mu\nu}\nabla_\mu\nabla_\nu)^k
	\end{equation}
	where $\phi$ is a placeholder used to represent all the fields present in the action under consideration and $\hat C(\phi)$ is an invertible matrix of coefficients corresponding to the $2k^{th}$ order kinetic terms in the action. In what follows, we briefly review the procedure to deal with non-minimal operators, which violate the condition of minimality (\ref{a9}).
	
	The condition of minimality (\hyperref[a9]{3.8}) was generalized to the condition of causality in Ref. \cite{nonminimal} to include models with degenerate kinetic terms among others,
	\begin{equation}\label{a11}
		\det \hat D(\nabla)= \det \hat C(\phi)\;.\; (g^{\mu\nu}\nabla_\mu\nabla_\nu)^m
	\end{equation}
	where $m=k \;\text{tr}\hat 1$ and `det' represents a functional determinant. Causality interpretation follows from the fact that the characteristic surface for these operators, defined by the equation $\det \hat D=0$, coincides with the light cone in the momentum space corresponding to each $\nabla_{\mu}$. 
    
    To deal with non-minimal operators, we consider a class of operators $\hat F (\nabla)$ such that (\ref{a10}) can be rewritten as,
	\begin{equation}\label{a14}
		\hat F(\nabla|\lambda)=\hat D(\nabla|\lambda)+\hat \Pi(\nabla|\lambda)
 	\end{equation}
 	where $\lambda$ is another parameter introduced to the system with a range $0\leq \lambda<\lambda_0$, where $\lambda_o$ is chosen such that the operator remains positive definite within the range. The operator $\hat F(\nabla|\lambda)$ is causal at all $\lambda$, and minimal at $\lambda=0$ (minimal gauge). This way the introduction of $\lambda$ helps us distinguish between the class of minimal and nonminimal operators constructed out of the general operator (\ref{a10}). The effective action for such an operator is given by,
	\begin{equation}\label{nonmineff}
		iW(\lambda)=iW(0)+\frac{1}{2}\int^\lambda_0 d\lambda' \text{Tr}\left[\frac{d\hat F(\lambda')}{d\lambda'}\hat G(\lambda')\right]
	\end{equation}
	where we have used the notation $\hat F(\lambda)$ to represent $\hat F(\nabla|\lambda)$. $W(0)$, here, represents the effective action for the minimal operator $\hat F(0)$, with a form similar to (\hyperref[eff]{2.4}), and $\hat G(\lambda)$ is the Green function corresponding to the non-minimal operator $\hat F(\lambda)$. Since the method for dealing with minimal operators is known, an operator $\hat K(\lambda)$ is chosen such that,
	\begin{equation}\label{kintro}
	    \hat F(\lambda)\hat K(\lambda) = \hat\Box^m + \hat M(\lambda)
	\end{equation}
	where $\hat D(\lambda)\hat K(\lambda)=\hat\Box^m +\hat K_1(\lambda)$ and $\hat \Pi(\lambda)\hat K(\lambda)+\hat K_1(\lambda)=\hat M(\lambda)$ where $\hat M(\lambda)$ is a $2m-1$ order differential operator. Note that with the introduction of $\hat K(\lambda)$, the operator $\hat F(\lambda)\hat K(\lambda)$ in (\ref{kintro}) is a minimal operator. The Green function is, then, obtained as an expansion in powers of perturbation $\hat M$:
	\begin{equation}
	\hat G = -\hat K\;.\;\sum_{n=0}^{\infty}(-1)^n\hat M_n\dfrac{\hat 1}{\hat\Box^{m(n+1)}}
	\end{equation}
    where
    \begin{equation}
   \hat M_0 = \hat M,\qquad\text{and}\qquad
   \hat M_{n+1} = \hat M\;.\;\hat M_n + [\hat \Box^n , \hat M_n] 
   \end{equation}

	One caveat of the method above is that the calculation of effective action becomes a bit tedious. As such the authors of Ref. \cite{nonminimal} have listed ways to simplify some of the steps by working out the calculations for some common theories. For a more detailed overview of the heat kernel approach in general, we would also recommend that the reader checks out Ref. \cite{scholar}. The paper \cite{toms} lists the consequences of choosing certain gauges for a theory, and also utilizes a method called the local momentum space technique \cite{effbook, bunch, yukawa}.
	
	Non-minimal operators relevant to quantum field theory and their corresponding heat kernel coefficients are listed in Ref. \cite{moss}. Throughout this paper, we will simply use (\hyperref[nonmineff]{3.11}) for our computations, without worrying about simplified methods, since they yield the same results \cite{nonminimal}.
	
    \section{Models under Consideration} \label{model}
    We first consider a simple scalar field model with $\phi^4$ self-interaction minimally coupled to gravity in Section \hyperref[modela1]{4.1.1} followed by adding a non-minimal coupling $\dfrac{1}{2}\xi\phi^2R$ in the action in Section \hyperref[modela2]{4.1.2}. The reason for considering the far simpler minimal-coupling case is two-fold: it serves as a good pedagogical example for the heat kernel technique, and it helps us establish the importance of Vilkovisky's prescription for the choice of field space metric as we go from the minimal to a nonminimal coupling case in Section \hyperref[modelb]{4.1.2} and later compare the results with Ref. \cite{mackay}. We shall also use \cite{stein} to see how the non-covariant results compare. In Section \hyperref[modelb]{4.2}, we consider a dilaton-extended two-field Starobinsky model in Einstein frame and verify the result obtained using the perturbative approach.
    
    Our purpose is to present a thorough calculation of only the divergent part of effective action at one-loop level. Throughout the paper, we shall assume that the spacetime background is flat and that the background scalar field is constant. In addition to making calculations easier, this limit is further motivated by its utilization when dealing with inflationary scenarios where it provides cosmologically relevant results despite its trivial nature (See Refs. \cite{gali,higg,ashish}).
    
	\subsection{Scalar Fields and Gravity} \label{modela}
	
	\subsubsection{Minimal Coupling}\label{modela1}
	We consider a model comprising of a single scalar field with $\phi^4$ self-interaction and Einstein-Hilbert gravitational action
	\begin{equation}\label{minimalaction}
	   S = \int d^4x \sqrt{g} \left(-\dfrac{2}{\kappa^2}R + \dfrac{1}{2}m^2\phi^2+\dfrac{1}{2}\partial_\mu\phi\partial^\mu\phi+\dfrac{\lambda\phi^4}{4!}\right)
	\end{equation}
	We adopt the following index notation: $g_{\alpha\beta} \rightarrow$ 1, $\phi \rightarrow$ 2. In the absence of a coupling term, the gravitational component of field-space metric corresponds to the Wheeler-DeWitt metric,
	\begin{equation}\label{gmetric}
	    G_{11}^{\mu\nu\alpha\beta}(x,x') = \dfrac{\sqrt{g(x)}}{2}[g^{\mu\alpha}(x)g^{\nu\beta}(x) + g^{\mu\beta}(x)g^{\nu\alpha}(x)  - g^{\mu\nu}(x)g^{\alpha\beta}(x) ] \tilde{\delta}(x,x')
	\end{equation}
	where $\tilde{\delta}(x,x')=\sqrt{g(x')}\delta(x,x')$.
	The scalar part of metric, as seen from the highest derivative terms of scalar field present in the action, is,
	\begin{equation}\label{phimetric}
	    G_{22}(x,x') = \sqrt{g(x)}\tilde{\delta}(x,x')
	\end{equation}
	To obtain the gauge fixing term we need to find the components $K^i_\alpha$ of vector \textbf{K}$_\alpha$ which generates gauge transformation (\hyperref[gtran]{2.9}). An infinitesimal change in the scalar field is given by
	\begin{equation}\label{delphi}
	    \delta\phi = -\delta\epsilon^\mu\partial_\mu\phi
	\end{equation}
	and under general coordinate transformations, change in spacetime metric field is given by,
	\begin{equation}\label{delg}
        \delta g_{\mu\nu} = -\delta\epsilon^\alpha\partial_\alpha g_{\mu\nu} - g_{\alpha\mu}\partial_\nu\epsilon^\alpha - g_{\alpha\nu}\partial_\mu\epsilon^\alpha
    \end{equation}
    Comparing (\hyperref[delphi]{4.4}) and (\hyperref[delg]{4.5}) with (\hyperref[gtran]{2.9}) gives,
    \begin{equation}
         K^1_{\mu\nu\,\alpha}(x,x') = -[\partial_\alpha g_{\mu\nu}(x) + g_{\alpha\mu}(x)\partial_\nu + g_{\alpha\nu}(x)\partial_\mu]\tilde{\delta}(x,x')
    \end{equation}
    and
    \begin{equation}
         K^2_\alpha(x,x') = -\partial_\alpha\phi(x)\tilde{\delta}(x,x')
    \end{equation}
    We choose Landau-DeWitt gauge so that the complicated Vilkovisky-DeWitt connections can be replaced by standard connections arising from field-space metric given in (\hyperref[gmetric]{4.2}) and (\hyperref[phimetric]{4.3}):
    \begin{equation}\label{ldgauge1}
        \chi_\alpha = K^i_\alpha[\bar{\varphi}]g_{ij}[\bar{\varphi}](\varphi^j - \bar{\varphi}^j)
    \end{equation}
    where $\bar{\varphi}^i$ are the background fields. The fields present in the action are split into background/classical fields and their quantum fluctuations as,
    \begin{equation}\label{qsplit}
        \begin{aligned}
            \phi &= \bar\phi + \delta\phi \\
            g_{\mu\nu} &= \eta_{\mu\nu} + \kappa h_{\mu\nu}
        \end{aligned}
    \end{equation}
    Using the expressions for field space metric components (\hyperref[gmetric]{4.2}) and (\hyperref[phimetric]{4.3}) in (\hyperref[ldgauge1]{4.8}) we get the gauge-fixing part of the action
    \begin{align}
        S_{GF} &= \dfrac{1}{2(1+\alpha)}\int d^4x \ \chi_\alpha(x)\chi^\alpha(x) \nonumber\\&=  \dfrac{1}{2(1+\alpha)}\int d^4x(2\partial^\mu g_{\mu\alpha} - \eta^{\mu\nu}\partial_\alpha g_{\mu\nu}-\phi\partial_\alpha\bar\phi)^2\nonumber\\&=\dfrac{1}{2(1+\alpha)}\int d^4x(2\partial^\mu h_{\mu\alpha} - \partial_\alpha h^\mu_\mu)^2\label{sgf}
    \end{align}
    where we have used (\hyperref[qsplit]{4.9}) and the fact that we are working around a background comprising flat spacetime and constant scalar field.\\
    The one-loop effective action is given by,
    \begin{equation}
        \Gamma^{(1)} =\lim_{\alpha \to -1}W(\alpha)= \lim_{\alpha \to -1}\dfrac{i}{2}\text{ln det}\left(\nabla_i\nabla^jS_{eff}[\alpha]\right)
    \end{equation}
    where $S_{eff}[\alpha] = S + S_{GF} + S_{Ghost}[\alpha]$. Here, $S_{Ghost}$ is the contribution from ghost term which comes from $Q_{\alpha\beta}$ given by:
    \begin{equation*}
        Q_{\alpha\beta} = \dfrac{\delta \chi_\alpha}{\delta \epsilon^\beta} = -2\eta_{\alpha\beta}\Box + \kappa^2\partial_\alpha\bar\phi\partial_\beta\bar\phi
    \end{equation*}
    For a constant scalar field background, it is clear that this expression won't contribute to the divergent part of effective action. To obtain first and ordinary second order derivatives, we expand the gauge-fixed action up to second order in quantum fields and then compute the coefficients of terms which are first and second order in quantum fields. The covariant derivative would then be obtained as,
    \begin{equation}\label{covder}
        \nabla_i\nabla_jS = S_{,ij} - \beta\Gamma^k_{ij}S_{,k}
    \end{equation}
    where a factor of $\beta$ is introduced to emphasize the presence of Vilkovisky-DeWitt connections (the motive shall become clear later on). Also for typographical convenience, we omit bars appearing over the background fields in all subsequent expressions. The connections arising from field-space metric are:
    \begin{align}
        ( \Gamma^1_{11})^{ \mu\nu\rho\sigma}_{\lambda\tau}(x,x',x'') =& \delta(x'',x')\delta(x'',x)\kappa\left(-\delta^{(\mu}_{(\lambda}g^{\nu)(\rho}\delta^{\sigma)}_{\tau)}+\dfrac{1}{4}g^{\mu\nu}\delta^{\rho}_{(\lambda}\delta^{\sigma}_{\tau)}\right.\nonumber+\dfrac{1}{4}g^{\rho\sigma}\delta^{\mu}_{(\lambda}\delta^{\nu}_{\tau)} \\&\qquad\qquad\qquad\qquad\qquad\qquad\left.+\dfrac{1}{4}g_{\lambda\tau}g^{\mu(\rho}g^{\sigma)\nu} -\dfrac{1}{8}g_{\lambda\tau}g^{\mu\nu}g^{\rho\sigma}\right)\nonumber\\
        ( \Gamma^1_{22})_{\mu\nu}(x,x',x'') =& \dfrac{1}{4}\tilde{\delta}(x'',x')\tilde{\delta}(x'',x)\kappa g_{\mu\nu}\nonumber\\
        (\Gamma^2_{12})^{\mu\nu}(x,x',x'') =&\dfrac{1}{4}\delta(x'',x')\tilde{\delta}(x'',x) \kappa g^{\mu\nu}
    \end{align}
    First and second order covariant derivatives with respect to the background fields are:
    \begin{align}\label{fd}
        \dfrac{\delta S_{eff}}{\delta g_{\mu\nu}} &= \sqrt{g}\left(\dfrac{1}{4}m^2\eta^{\mu\nu}\phi^2 +\dfrac{1}{48}\lambda\eta^{\mu\nu}\phi^4\right)\\
        \dfrac{\delta S_{eff}}{\delta \phi} &=\sqrt{g}\left( m^2\phi +\dfrac{1}{6}\lambda\phi^3\right)
    \end{align}
    Note that since we have a background consisting of flat spacetime and constant scalar field the Einstein-Hilbert term and kinetic term of scalar field, do not contribute to (\hyperref[fd]{4.14, 4.15}).
    \begin{multline}\label{seffgg}
            S_{eff;g_{\mu\nu}(x)g_{\rho\sigma}(x')} =\sqrt{g}\tilde{\delta}(x,x')\Bigg[ \left(\dfrac{1}{1+\alpha}-1\right)\Big(\eta^{\mu\nu}\nabla^\rho\nabla^\sigma + \eta^{\rho\sigma}\nabla^\mu\nabla^\nu - \dfrac{1}{2}\eta^{\nu\rho}\nabla^\mu\nabla^\sigma\\-\dfrac{1}{2}\eta^{\nu\sigma}\nabla^\mu\nabla^\rho-\dfrac{1}{2}\eta^{\mu\rho}\nabla^\nu\nabla^\sigma-\dfrac{1}{2}\eta^{\mu\sigma}\nabla^\nu\nabla^\rho\Big) - \dfrac{1}{2}(\eta^{\mu\sigma}\eta^{\nu\rho} + \eta^{\mu\rho}\eta^{\nu\sigma} - \eta^{\mu\nu}\eta^{\rho\sigma})\Box \\+ \dfrac{1}{2}\left(1-\dfrac{1}{1+\alpha}\right)\eta^{\mu\nu}\eta^{\rho\sigma}\Box -\left(\dfrac{\lambda\phi^2}{96}+\dfrac{m^2}{8}\right)\kappa^2\phi^2(\eta^{\mu\sigma}\eta^{\nu\rho} + \eta^{\mu\rho}\eta^{\nu\sigma} - \eta^{\mu\nu}\eta^{\rho\sigma})\Bigg]
    \end{multline}
    (Note the absence of $\beta$ in this expression.)
    \begin{equation}\label{seffphiphi}
        S_{eff;\phi(x)\phi(x')} =\sqrt{g}\tilde{\delta}(x,x')\Bigg[ -\Box + m^2 + \dfrac{1}{2}\lambda\phi^2 \\- \dfrac{1}{4}\beta\kappa^2m^2\phi^2 - \dfrac{1}{48}\beta\kappa^2\lambda\phi^4\Bigg]
    \end{equation}
    \begin{equation}\label{seffgphi}
        S_{eff;g_{\mu\nu}(x)\phi(x)} = \sqrt{g}\tilde{\delta}(x,x')\Bigg[-\dfrac{1}{4}\beta\kappa m^2\phi \ \eta^{\mu\nu} -\dfrac{1}{24}\beta\kappa\lambda\phi^3\eta^{\mu\nu} +\dfrac{1}{2}\kappa m^2\phi \ \eta^{\mu\nu} + \dfrac{1}{12}\kappa\lambda\phi^3\eta^{\mu\nu}\Bigg]
    \end{equation}
    The operator $\nabla_i\nabla_kS_{eff}$ must be multiplied with the inverse of field-space metric $G^{jk}$ to get $\nabla_i\nabla^jS_{eff}$:
    \begin{equation}\label{nonminop}
        \nabla_i\nabla^jS_{eff} = \sqrt{g}\tilde\delta(x,x')(-\delta_i^{ \ j}\Box + A_i^{ \ j}(\nabla) + Q_i^{ \ j})
    \end{equation}
    where
    \begin{equation}
        \delta_i^{ \ j} = \begin{pmatrix} \dfrac{1}{2}(\delta_\mu^{\ \rho}\delta_\nu^{ \ \sigma} + \delta_\mu^{\ \sigma}\delta_\nu^{\ \rho}) & \hspace{4mm}0 \\ 0 & \hspace{4mm}1 \nonumber
        \end{pmatrix}
    \end{equation}
    \begin{equation}\nonumber
        A_i^{\ j}(\nabla) = \begin{pmatrix}\dfrac{1}{2}\Bigg(1-\dfrac{1}{1+\alpha}\Bigg)(\delta^{(\rho}_{(\mu}\nabla^{\sigma)}\nabla_{\nu)}-2\eta^{\rho\sigma}\nabla_\mu\nabla_\nu) &\hspace{4mm} 0 \\ 0 &\hspace{4mm} 0\end{pmatrix}
    \end{equation}
    
    \begin{equation}
        Q_i^{\ j} = \begin{pmatrix}-\left(m^2+\dfrac{\lambda\phi^2}{12}\right)\dfrac{\kappa^2\phi^2}{8}(\delta_\mu^{ \ \rho}\delta_\nu^{\ \sigma} + \delta_\mu^{\ \sigma}\delta_\nu^{\ \rho})\quad & \left(m^2-\dfrac{\beta m^2}{2} -\dfrac{\beta\lambda\phi^2}{12} + \dfrac{\lambda\phi^2}{6}\right)\dfrac{\kappa\phi\eta^{\rho\sigma}}{2} \\ \left(-m^2 - \dfrac{\lambda\phi^2}{6}+\dfrac{\beta m^2}{2} +\dfrac{\beta\lambda\phi^2}{12}\right)\dfrac{\kappa\phi\eta^{\mu\nu}}{2}\quad &  m^2 + \dfrac{\lambda\phi^2}{2} - \dfrac{\beta\kappa^2m^2\phi^2}{4} - \dfrac{\beta\kappa^2\lambda\phi^4}{48} \end{pmatrix}
    \end{equation}
    \noindent Note that the Latin indices ($i, j, k,..$) have been used throughout this paper as labels for the components of the field space, and as such they contain information like the spacetime indices of the corresponding field. It is readily seen from (\hyperref[seffgg]{4.16}), (\hyperref[seffphiphi]{4.17}), and (\hyperref[seffgphi]{4.18}) that $\alpha = 0$ reduces the differential operator $\nabla_i\nabla_kS_{eff}$ to a minimal form. A gauge with such a choice of gauge parameter $\alpha$ is also called the minimal gauge. Keeping effective action up to quartic order in background fields, we get the first term on R.H.S. of (\hyperref[nonmineff]{3.11})  corresponding to minimal gauge to be,
    \begin{multline}
            W(0) = \frac{1}{16\pi^2\epsilon}\int d^4x\Bigg(-\dfrac{1}{2}m^4+\kappa^2m^4\phi^2 - \dfrac{3}{4}\beta\kappa^2m^4\phi^2 + \dfrac{1}{4}\beta^2\kappa^2m^4\phi^2 -\dfrac{1}{2}m^2\phi^2\lambda\\+ \dfrac{1}{3}m^2\kappa^2\lambda\phi^4 - \dfrac{3}{16}m^2\beta\kappa^2\lambda\phi^4+\dfrac{1}{12}m^2\beta^2\kappa^2\lambda\phi^4 - \dfrac{1}{8}\lambda^2\phi^4\Bigg)
    \end{multline}
    where $\epsilon = 4 - n$. Taking $\beta = 1$ we get,
    \begin{equation}
     W(0) = \frac{1}{16\pi^2\epsilon}\int d^4x\Bigg(-\dfrac{1}{2}m^4 + \dfrac{1}{2}\kappa^2m^4\phi^2 -\dfrac{1}{2}m^2\phi^2\lambda+\dfrac{11}{48}m^2\kappa^2\lambda\phi^4 -\dfrac{1}{8}\lambda^2\phi^4\Bigg)\label{w0}
     \end{equation}
    If instead of (\hyperref[covder]{4.12}), we had only used ordinary derivatives, which amounts to taking $\beta = 0$, we would get the non-covariant result:
    \begin{equation}
        W(0) = \frac{1}{16\pi^2\epsilon}\int d^4x \Bigg(-\dfrac{1}{2}m^4 -\dfrac{1}{2}m^2\phi^2\lambda + \kappa^2m^4\phi^2+\dfrac{1}{3}m^2\kappa^2\lambda\phi^4 - \dfrac{1}{8}\lambda^2\phi^4\Bigg)\label{excconnec}
    \end{equation}
    To proceed with our calculations, we now follow the procedure elaborated in Ref. \cite{nonminimal} to obtain the second term in (\hyperref[nonmineff]{3.11}). The choice of $\hat K(\nabla)$ which brings operator  (\hyperref[nonminop]{4.19}) to minimal form is:
    \begin{equation}
        \hat K(\nabla) = \begin{pmatrix}\delta^\alpha_{(\mu}\delta^\beta_{\nu)}\Box+2\alpha\delta^{(\alpha}_{(\mu}\nabla^{\beta)}\nabla_{\nu)}-\alpha\eta^{\alpha\beta}\nabla_{\mu}\nabla_{\nu} & 0 \\ 0 & \Box \end{pmatrix}
    \end{equation}
    After some rudimentary calculations, contribution from the second term in (\hyperref[nonmineff]{3.11}) is found to be,
    \begin{equation}
    \dfrac{i}{2}\int_0^\alpha d\alpha' \Bigg(\dfrac{m^4\kappa^2\phi^2}{8} + \dfrac{m^2\kappa^2\lambda\phi^4}{24}\Bigg) \text{Tr}\left(\dfrac{1}{\Box^2}\delta(x,x')\right)\Bigg|^{x = x'}_{div}= -\dfrac{\alpha m^4\kappa^2\phi^2}{128\pi^2\epsilon} - \dfrac{\alpha m^2\kappa^2\lambda\phi^4}{384\pi^2\epsilon}\label{walpha}
    \end{equation}
     We now set $\beta=1$ to obtain a covariant result. Combining (\hyperref[w0]{4.22}) and (\hyperref[walpha]{4.25}) we get,
    \begin{multline}
        W(\alpha) = \frac{1}{16\pi^2\epsilon}\int d^4x \Bigg(-\dfrac{1}{2}m^4 -\dfrac{1}{2}m^2\phi^2\lambda + \dfrac{1}{2}\kappa^2m^4\phi^2+\dfrac{11}{48}m^2\kappa^2\lambda\phi^4 - \dfrac{1}{8}\lambda^2\phi^4\\-\dfrac{1}{8}\alpha m^4\kappa^2\phi^2 -\dfrac{1}{24}\alpha m^2\kappa^2\lambda\phi^4 \Bigg)
    \end{multline}
    We can now safely take $\alpha \rightarrow -1$, since there are no terms that blow up in that limit.
    \begin{equation}
        \Gamma^{(1)}_{div} =\frac{1}{16\pi^2\epsilon}\int d^4x\Bigg( -\dfrac{1}{2}m^4 -\dfrac{1}{2}m^2\phi^2\lambda + \dfrac{5}{8}\kappa^2m^4\phi^2+\dfrac{13}{48}m^2\kappa^2\lambda\phi^4 - \dfrac{1}{8}\lambda^2\phi^4\Bigg)\label{gam1div1}
    \end{equation}

	\subsubsection{Non-minimal Coupling}\label{modela2}
    We now add a coupling term to the theory above,
    \begin{equation}\label{nonminaction}
    S = \int d^4x \sqrt{g} \Biggl(-\dfrac{2}{\kappa^2}R +\dfrac{1}{2}\xi\phi^2R + \dfrac{1}{2}m^2\phi^2+\dfrac{1}{2}\partial_\mu\phi\partial^\mu\phi+\dfrac{1}{4!}\lambda\phi^4\Biggr)
    \end{equation}
	The field-space metric may not be apparent from the action above. We proceed by considering functions $F[\phi]$, $J[\phi]$ and $H[\phi]$ such that,
	\begin{align}
	    G_{11}^{\mu\nu\alpha\beta}(x,x') &= \dfrac{\sqrt{g(x)}}{2}F[\phi]\Bigl(g^{\mu\alpha}(x)g^{\nu\beta}(x) + g^{\mu\beta}(x)g^{\nu\alpha}(x) - g^{\mu\nu}(x)g^{\alpha\beta}(x)\Bigr)\tilde{\delta}(x,x')\\
	    G_{12}^{\mu\nu}(x,x') &= \sqrt{g(x)}H[\phi]g^{\mu\nu}(x)\tilde{\delta}(x,x')\\G_{22}(x,x')&=\sqrt{g(x)}J[\phi]\tilde{\delta}(x,x')
	\end{align}
	Evaluating $K^i_\lambda g_{ij}(\varphi^j-\bar\varphi^j)$:
	\begin{align}
	\int d^4x' d^4x'' K^1_{\mu\nu\alpha}(x,x')G_{11}^{\mu\nu\rho\sigma}(x',x'')(g_{\rho\sigma}(x'')-\eta_{\rho\sigma}(x''))
	&=F[\bar\phi](2\partial^\mu g_{\mu\alpha}-\eta^{\mu\nu}\partial_{\alpha}g_{\mu\nu})\\
	\int d^4x' d^4x'' K^1_{\mu\nu\alpha}(x,x')G_{12}^{\mu\nu}(x',x'')(\phi(x'')-\bar\phi(x''))&=2H[\bar\phi]\partial_\alpha \phi
	\end{align}
	Thus, we find:
	\begin{equation}\nonumber
	    \chi_\alpha = F[\bar\phi](2\partial^\mu g_{\mu\alpha}-\eta^{\mu\nu}\partial_{\alpha}g_{\mu\nu}) +2H[\bar\phi]\partial_\alpha \phi 
	\end{equation}
	For convenience, we put a factor of $1/\sqrt{F[\phi]}$ in $\chi_\alpha$ that doesn't alter the gauge.
	\begin{equation}\label{nonmingauge}
	      \chi_\alpha = \dfrac{F[\bar\phi](-2\partial^\mu g_{\mu\alpha}+\eta^{\mu\nu}\partial_{\alpha}g_{\mu\nu}) -2H[\bar\phi]\partial_\alpha \phi}{\sqrt{F[\bar\phi]}}
	\end{equation}
The gauge-fixing action reads:
	\begin{equation}
	    S_{GF} = \int d^4x \dfrac{1}{F[\bar\phi]}(F[\bar\phi](2\partial^\mu h_{\mu\alpha} - \partial_\alpha h)+2H[\bar\phi]\partial_\alpha\delta\phi)^2
	\end{equation}
	Using (\hyperref[nonmingauge]{4.34}), we find,
	\begin{equation*}
	    Q_{\alpha\beta} = \dfrac{-2F[\bar\phi]\eta_{\alpha\beta}\Box - 2H[\bar\phi]\partial_\alpha\partial_\beta\bar\phi}{\sqrt{F[\bar\phi]}}
	\end{equation*}
	For a constant scalar field background, the above term does not contribute to the divergent part of effective action. We find the second order ordinary derivatives around the background to be (again omitting bars over background fields):
	\begin{multline}\label{seffgg1}
	    S_{eff,g_{\mu\nu}(x) g_{\rho\sigma}(x')}=\sqrt{g}\tilde{\delta}(x,x')\Bigg[ \left(\dfrac{F[\phi]}{1+\alpha}+\dfrac{1}{4}\kappa^2\xi\phi^2-1\right)\Biggl(\eta^{\mu\nu}\nabla^\rho\nabla^\sigma + \eta^{\rho\sigma}\nabla^\mu\nabla^\nu - \dfrac{1}{2}\eta^{\nu\rho}\nabla^\mu\nabla^\sigma\\-\dfrac{1}{2}\eta^{\nu\sigma}\nabla^\mu\nabla^\rho-\dfrac{1}{2}\eta^{\mu\rho}\nabla^\nu\nabla^\sigma-\dfrac{1}{2}\eta^{\mu\sigma}\nabla^\nu\nabla^\rho\Biggr) - \dfrac{1}{2}\Bigg[\Biggl(1-\dfrac{1}{4}\kappa^2\xi\phi^2\Biggr)(2\eta^{\mu(\sigma}\eta^{\rho)\nu}  -\eta^{\mu\nu}\eta^{\rho\sigma}) \\- \Bigg(1 - \dfrac{F[\phi]}{1+\alpha}-\dfrac{1}{4}\kappa^2\xi\phi^2\Bigg) \eta^{\mu\nu}\eta^{\rho\sigma}\Bigg]\Box - \Bigg(\dfrac{\lambda\phi^2}{96} +\dfrac{m^2}{8}\Bigg)\kappa^2\phi^2(2\eta^{\mu(\sigma}\eta^{\rho)\nu} - \eta^{\mu\nu}\eta^{\rho\sigma})\Bigg]
	\end{multline}
	\begin{multline}\label{seffgphi1}
	        S_{eff,g_{\mu\nu}(x)\phi(x')} =\sqrt{g}\tilde{\delta}(x,x')\Biggl[2\Biggl(-\dfrac{H[\phi]}{1+\alpha} +\dfrac{\kappa\xi\phi}{2}\Biggr)\nabla^\mu\nabla^\nu  -\Biggl(\kappa\xi\phi -\dfrac{H[\phi]}{1+\alpha}\Biggr)\eta^{\mu\nu}\Box \\+\dfrac{1}{2}\Biggl(m^2\kappa\xi+\dfrac{\kappa\lambda\phi^2}{6}\Biggr)\eta^{\mu\nu}\phi\Biggr]
	\end{multline}
	\begin{equation}\label{seffphiphi1}
	        S_{eff,\phi(x)\phi(x')}= \sqrt{g}\tilde{\delta}(x,x')\Biggl[\Biggl(-1-\dfrac{2H^2[\phi]}{(1+\alpha)F[\phi]}\Biggr)\Box +m^2 +\dfrac{\lambda\phi^2}{2} \Biggr]
	\end{equation}
	From (\hyperref[seffgg1]{4.36}),(\hyperref[seffgphi1]{4.37}), and (\hyperref[seffphiphi1]{4.38}), it is seen that the choice of functions that would turn $S_{eff,ij}$ into minimal form when $\alpha$ is taken zero is,
    \begin{equation}
        \begin{aligned}
            F[\phi] = 1-\dfrac{1}{4}\kappa^2\xi\phi^2, \ \; \ \ 
            H[\phi] = \dfrac{1}{2}\kappa\xi\phi
        \end{aligned}
    \end{equation}
    Using $F[\phi]$ and $H[\phi]$ in (\hyperref[seffphiphi1]{4.38}) we find:
    \begin{equation}
        G_{22}(x,x') = \sqrt{g(x)}\left(1 + \dfrac{\dfrac{1}{2}\kappa^2\xi^2\phi^2(x)}{1-\dfrac{1}{4}\kappa^2\xi\phi^2(x)}\right)\tilde{\delta}(x,x')
    \end{equation}
    so that,
    \begin{equation}
        J[\phi] = \left(1 + \dfrac{\dfrac{1}{2}\kappa^2\xi^2\phi^2(x)}{1-\dfrac{1}{4}\kappa^2\xi\phi^2(x)}\right)
    \end{equation}
    Since we intend to find effective action up to second order in $\kappa$, we find our field-space metric up to $\mathcal{O}(\kappa ^2)$ to be:
    \begin{equation}\label{nonminmetric}
        G_{ij} = \sqrt{g}\tilde\delta(x,x')\begin{pmatrix}\Biggl(1-\dfrac{1}{4}\kappa^2\xi\phi^2\Biggr)\mathcal{G}^{\mu\nu\alpha\beta} & \dfrac{1}{2}\kappa\xi\phi g^{\mu\nu}\\
    \dfrac{1}{2}\kappa\xi\phi g^{\alpha\beta} & 1+\dfrac{1}{2}\kappa^2\xi^2\phi^2\end{pmatrix}
    \end{equation}
    where $\mathcal{G}^{\mu\nu\alpha\beta}$ is the Wheeler-DeWitt metric:
    \begin{equation}\label{wdmetric}
        \mathcal{G}^{\mu\nu\alpha\beta} = \dfrac{1}{2}(g^{\mu\alpha}g^{\nu\beta} + g^{\mu\beta}g^{\nu\alpha}-g^{\mu\nu}g^{\alpha\beta})
    \end{equation}
	The inverse field-space metric is found to be:
	\begin{equation}
	    G^{ij} = \delta(x,x')\begin{pmatrix}\Biggl(1+\dfrac{1}{4}\kappa^2\xi\phi^2\Biggr)\mathcal{G}_{\mu\nu\alpha\beta} & \dfrac{1}{2}\kappa\xi\phi g_{\mu\nu}\\
    \dfrac{1}{2}\kappa\xi\phi g_{\alpha\beta} & 1-\dfrac{3}{2}\kappa^2\xi^2\phi^2\end{pmatrix}
	\end{equation}
	The connections arising out of the field-space metric in (\hyperref[nonminmetric]{4.42}) are:
	
	\begin{align}
        (\Gamma^1_{11})^{ \mu\nu\rho\sigma}_{\lambda\tau}(x,x',x'') =& \tilde{\delta}(x'',x')\tilde{\delta}(x'',x)\kappa\left(-\delta^{(\mu}_{(\lambda}g^{\nu)(\rho}\delta^{\sigma)}_{\tau)} +\dfrac{1}{4}g^{\mu\nu}\delta^{\rho}_{(\lambda}\delta^{\sigma}_{\tau)}+\dfrac{1}{4}g^{\rho\sigma}\delta^{\mu}_{(\lambda}\delta^{\nu}_{\tau)}\right.\nonumber\\&\qquad\qquad\qquad\qquad\qquad\qquad\quad +\left.\dfrac{1}{4}g_{\lambda\tau}g^{\mu(\rho}g^{\sigma)\nu} -\dfrac{1}{8}g_{\lambda\tau}g^{\mu\nu}g^{\rho\sigma}\right) \nonumber\\
	   (\Gamma^1_{12})^{\mu\nu}_{\rho\sigma}(x,x',x'') =& \tilde{\delta}(x'',x')\tilde{\delta}(x'',x)\Bigl[\dfrac{\kappa^2\xi\phi}{8}(g_{\rho\sigma}g^{\mu\nu} - 2\delta^\mu_{(\rho}\delta^\nu_{\sigma)})\Bigr]\nonumber\\
	   (\Gamma^1_{22})_{\mu\nu}(x,x',x'') =&\tilde{\delta}(x'',x')\tilde{\delta}(x'',x)\Bigl[\dfrac{\kappa g_{\mu\nu}}{4}(1-2\xi)\Bigr] \nonumber\\
	   (\Gamma^2_{11})^{\mu\nu\rho\sigma}(x,x',x'')=&\tilde{\delta}(x'',x')\tilde{\delta}(x'',x)\Bigr[\dfrac{\kappa^2\xi\phi}{8}(g^{\mu\nu}g^{\rho\sigma}-2g^{\mu(\rho}g^{\sigma)\nu})\Bigr]\nonumber\\
	   (\Gamma^2_{12})^{\mu\nu}(x,x',x'')=&\dfrac{1}{4}\tilde{\delta}(x'',x')\tilde{\delta}(x'',x)\kappa g^{\mu\nu} \nonumber\\
	   (\Gamma^2_{22})(x,x',x'')=&\tilde{\delta}(x'',x')\tilde{\delta}(x'',x)\left(\dfrac{3}{2}\kappa^2\xi^2\phi-\dfrac{1}{2}\kappa^2\xi\phi\right)
	\end{align}
    For $\xi = 0$, the field-space metric and the corresponding connections are seen to correspond to the case of minimal coupling as seen in Section  \hyperref[modela1]{4.1.1}. The connections are used to calculate $\nabla_i\nabla_jS_{eff}$ 
    around background (again omitting bars over background fields):
	\begin{multline}
	    S_{eff;g_{\mu\nu}(x) g_{\rho\sigma}(x')}=\sqrt{g}\tilde{\delta}(x,x')\Bigg[ \left(\dfrac{F[\phi]}{1+\alpha}+\dfrac{1}{4}\kappa^2\xi\phi^2-1\right)\Biggl(\eta^{\mu\nu}\nabla^\rho\nabla^\sigma + \eta^{\rho\sigma}\nabla^\mu\nabla^\nu - \dfrac{1}{2}\eta^{\nu\rho}\nabla^\mu\nabla^\sigma\\-\dfrac{1}{2}\eta^{\nu\sigma}\nabla^\mu\nabla^\rho-\dfrac{1}{2}\eta^{\mu\rho}\nabla^\nu\nabla^\sigma-\dfrac{1}{2}\eta^{\mu\sigma}\nabla^\nu\nabla^\rho\Biggr) - \dfrac{1}{2}\Bigg[\Biggl(1-\dfrac{1}{4}\kappa^2\xi\phi^2\Biggr)(2\eta^{\mu(\sigma}\eta^{\rho)\nu} -\eta^{\mu\nu}\eta^{\rho\sigma})\\ - \Bigg(1 - \dfrac{F[\phi]}{1+\alpha}-\dfrac{1}{4}\kappa^2\xi\phi^2\Bigg) \eta^{\mu\nu}\eta^{\rho\sigma}\Bigg]\Box - \Bigg(\dfrac{\lambda\phi^2}{96}\Big(1-2\beta\xi\Big)+\dfrac{m^2}{8}\Big(1-\beta\xi\Big)\Bigg)\kappa^2\phi^2(2\eta^{\mu(\sigma}\eta^{\rho)\nu} - \eta^{\mu\nu}\eta^{\rho\sigma})\Bigg]
	\end{multline}
	\begin{multline}\label{seffcgphi1}
        S_{eff;g_{\mu\nu}(x)\phi(x')} =\sqrt{g}\tilde{\delta}(x,x')\Biggl[2\Biggl(-\dfrac{H[\phi]}{1+\alpha} +\dfrac{\kappa\xi\phi}{2}\Biggr)\nabla^\mu\nabla^\nu  -\Biggl(\kappa\xi\phi -\dfrac{H[\phi]}{1+\alpha}\Biggr)\eta^{\mu\nu}\Box \\+\dfrac{1}{2}\Biggl(m^2\kappa\xi+\dfrac{\kappa\lambda\phi^2}{6}\Biggr)\Bigl(1 - \dfrac{\beta}{2}\Bigr)\eta^{\mu\nu}\phi\Biggr]
	\end{multline}
	\begin{multline}\label{seffcphiphi1}
	        S_{eff;\phi(x)\phi(x')}= \sqrt{g}\tilde{\delta}(x,x')\Biggl[\Biggl(-1-\dfrac{2H^2[\phi]}{(1+\alpha)F[\phi]}\Biggr)\Box +m^2 +\dfrac{\lambda\phi^2}{2} -\dfrac{1}{4}\beta\kappa^2m^2\phi^2 +\beta\kappa^2 m^2\xi\phi^2 \\- \dfrac{3}{2}\beta\kappa^2m^2\xi^2\phi^2- \dfrac{1}{48}\beta\kappa^2\lambda\phi^4 +\dfrac{1}{8}\beta\kappa^2\lambda\xi\phi^4-\dfrac{1}{4}\beta\kappa^2\lambda\xi^2\phi^4\Biggr]
	\end{multline}
	which is then multiplied by inverse field-space metric to give $\nabla_i\nabla^jS_{eff}$. Corresponding to the minimal gauge,
    \begin{multline}
        W(0) = \frac{1}{16\pi^2\epsilon}\int d^4x\Bigg(-\dfrac{1}{2}m^4 + \kappa^2m^4\phi^2 - \dfrac{3}{4}\beta\kappa^2m^4\phi^2 + \dfrac{1}{4}\beta^2\kappa^2m^4\phi^2  -\dfrac{1}{2}m^2\phi^2\lambda   -2m^4\kappa^2\xi\phi^2 \\+ \dfrac{3}{2}m^4\kappa^2\xi^2\phi^2+ \dfrac{3}{2}m^4\beta\kappa^2\xi^2\phi^2 + \dfrac{1}{3}m^2\kappa^2\lambda\phi^4 - \dfrac{3}{16}m^2\beta\kappa^2\lambda\phi^4+\dfrac{1}{12}m^2\beta^2\kappa^2\lambda\phi^4 \\- \dfrac{1}{8}\lambda^2\phi^4-\dfrac{4}{3}m^2\kappa^2\lambda\xi\phi^4 + \dfrac{1}{24}m^2\beta\kappa^2\lambda\xi\phi^4+\dfrac{3}{2}m^2\kappa^2\lambda\xi^2\phi^4 + m^2\beta\kappa^2\lambda\xi^2\phi^4\Bigg)
    \end{multline}
    Taking $\beta = 1$, we get:
    \begin{multline}
        W(0) = \frac{1}{16\pi^2\epsilon}\int d^4x \Bigg(-\dfrac{1}{2}m^4 + \dfrac{1}{2}\kappa^2m^4\phi^2 -\dfrac{1}{2}m^2\phi^2\lambda  -2m^4\kappa^2\xi\phi^2 + 3m^4\kappa^2\xi^2\phi^2 + \dfrac{11}{48}m^2\kappa^2\lambda\phi^4 \\-\dfrac{1}{8}\lambda^2\phi^4 - \dfrac{31}{24}m^2\kappa^2\lambda\xi\phi^4 + \dfrac{5}{2}m^2\kappa^2\lambda\xi^2\phi^4\Bigg)
    \end{multline}
    If we exclude the connection terms by taking $\beta = 0$, then, we get:
    \begin{multline}\label{nonminbetazero}
        W(0) = \frac{1}{16\pi^2\epsilon}\int d^4x\Bigg(
        -\dfrac{1}{2}m^4 + \kappa^2m^4\phi^2 -\dfrac{1}{2}m^2\phi^2\lambda   -2m^4\kappa^2\xi\phi^2 + \dfrac{3}{2}m^4\kappa^2\xi^2\phi^2\\ + \dfrac{1}{3}m^2\lambda\kappa^2\phi^4 - \dfrac{1}{8}\lambda^2\phi^4 - \dfrac{4}{3}m^2\kappa^2\lambda\xi\phi^4 + \dfrac{3}{2}m^2\kappa^2\lambda\xi^2\phi^4\Bigg)
    \end{multline}
    The operator $\hat K(\nabla)$, which brings operator $\nabla_i\nabla^jS_{eff}$ to a minimal form, is found to be:
    \begin{equation}
        \hat K(\nabla) = \begin{pmatrix}\delta^\alpha_{(\mu}\delta^\beta_{\nu)}\Box+2\alpha\delta^{(\alpha}_{(\mu}\nabla^{\beta)}\nabla_{\nu)}-\alpha\eta^{\alpha\beta}\nabla_{\mu}\nabla_{\nu} & 0 \\ \alpha\kappa\xi\phi\nabla_\mu\nabla_\nu & \Box \end{pmatrix}
    \end{equation}
    For the second term of (\hyperref[nonmineff]{3.11}), we encounter the same universal trace (\hyperref[walpha]{4.25}) that we found in the previous Section. Again, taking the limit $\alpha \rightarrow -1$, we get:
    \begin{multline}\label{gam1div2}
            \Gamma^{(1)}_{div}  = \frac{1}{16\pi^2\epsilon}\int d^4x \Bigg(-\dfrac{1}{2}m^4 -\dfrac{1}{2}m^2\phi^2\lambda + \dfrac{5}{8}\kappa^2m^4\phi^2 -2m^4\kappa^2\xi\phi^2 + 3m^4\kappa^2\xi^2\phi^2\\+ \dfrac{13}{48}m^2\kappa^2\lambda\phi^4 -\dfrac{1}{8}\lambda^2\phi^4 - \dfrac{31}{24}m^2\kappa^2\lambda\xi\phi^4 + \dfrac{5}{2}m^2\kappa^2\lambda\xi^2\phi^4\Bigg)
    \end{multline}

	\subsection{Starobinsky Model with additional Scalar Field}\label{modelb}
    It is well established that a brief period of inflation resolves several problems that plagued the initial versions of Big Bang cosmology. Several models of inflation have been devised whose predictions of spectral tilt $n_s$ and tensor to scalar ratio $r$ agree well with the experimental CMB data. Standard inflation models, involving a single scalar field, are able to explain much of the current cosmological data. A natural generalisation to this is a multi-field inflation model, which provides new predictions like isocurvature perturbations, in addition to adiabatic perturbations present in single-field inflation.
    
    In most inflationary models, the driving force of inflation is the false vacuum energy density $\rho$ which leads to a de Sitter exponential expansion $\exp(Ht)$, where $H = \sqrt{(8\pi G)\rho/3}$, called the Hubble's constant.
    
    Starobinsky's model \cite{staro} is different in this aspect. A key feature of this model is that exponential expansion of the universe is brought about naturally as a de Sitter solution to Einstein's equations, where one-loop quantum corrections are taken into account. Such a solution is unstable as first pointed out by Starobinsky and leads to a hot Friedmann universe. In this way, unlike old inflation models, Starobinsky's model never faced the graceful exit problem. Starobinsky noted that an $R^2$ term in the action played an important role in a large curvature regime. Such a modification to Einstein's gravity leads to an effective cosmological constant to drive the inflation. The origin of $R^2$ term can be explained roughly as follows: starting with the Einstein-Hilbert action, when we work out the one-loop corrections, we arrive at divergent quantities of the form $R_{\mu\nu} R^{\mu\nu}$ and $R^2$. For a de Sitter space, $R_{\mu\nu} = \dfrac{1}{4}g_{\mu\nu}R$ and $\nabla_\mu R = 0$. This implies that on-shell corrections only bring in $R^2$-divergent terms, which explains the origin of $R^2$ term in Starobinsky's action:
    \begin{equation}\label{starob}
        S = \dfrac{2}{\kappa^2}\int d^4x \sqrt{g}\Biggl(R + \dfrac{R^2}{6m^2}\Biggr)
    \end{equation}
    where the parameter $m$ is experimentally determined from CMB amplitude observations. The $R^2$ term bears the interpretation of an additional degree of freedom, and is often labeled as the `scalaron' and assigned mass $m$, bringing forth a scalar-tensor representation of the aforementioned action.
    
    In this Section, we consider a two-field Starobinsky-like model where a dilaton field $\phi$ is non-minimally coupled to the action in (\hyperref[starob]{4.54}), as detailed in Refs. \cite{salvio,dilaton,iaon2}. The modified action is given by,
    \begin{equation}
        S = \int d^4x \sqrt{g}\Biggl[ -\dfrac{U(\phi)}{2}\Biggl(R+\dfrac{R^2}{6M^2(\phi)}\Biggr)+V(\phi)+\dfrac{1}{2}\partial_\mu\phi\partial^\mu\phi\Biggr]
    \end{equation}
	with 
	\begin{equation}
	        V(\phi) = \dfrac{1}{4}\phi^4, \qquad U(\phi) = \dfrac{4}{\kappa^2} + \xi\phi^2, \qquad M^2(\phi) = \dfrac{\kappa^2}{4}(m_o^2 + \zeta\phi^2)U(\phi)
	\end{equation}
	where $m_o$ is the scalaron mass.
    The action here is defined in the Jordan frame and is transformed to Einstein frame through a conformal transformation and field redefinition using the procedure followed in Ref. \cite{JtoE}. The Einstein frame action, on which we would be performing our calculations, is given by,
    \begin{equation}\label{actionb}
        S = \int d^4x\sqrt{g}\Biggl(-\dfrac{2}{\kappa^2}R +\dfrac{1}{2}G_{IJ}\partial_\mu\Phi^I\partial^\mu\Phi^J+W(\chi,\phi)\Biggr)
    \end{equation}
	where
	\begin{equation}
        \Phi^I = \begin{pmatrix}\chi \\ \phi \end{pmatrix} ,\text{ \ \ \ \ \ \ \ }G_{IJ} = \begin{pmatrix} 1 & 0 \\ 0 & F^{-1} \end{pmatrix}, \nonumber
	\end{equation}
	\begin{equation}
	        W(\chi,\phi) = \dfrac{1}{4F^2}\Biggl[\lambda\phi^4 + \frac{12}{\kappa^2}\bigl(m_o^2 + \zeta\phi^2\bigr)\Bigl(1+\frac{\kappa^2}{4}\xi\phi^2 - F\Bigr)^2\Biggr],\quad\text{and}\quad F(\chi) = \text{exp}\left(\dfrac{\kappa\chi}{\sqrt{6}}\right)
	\end{equation}
    The notation used here is: $g_{\alpha\beta} \rightarrow 1$, $\chi \rightarrow 2$, $\phi \rightarrow 3$. The field-space metric is simply,
	\begin{align}
	   G_{11}^{\mu\nu\alpha\beta}(x,x') &= \dfrac{\sqrt{g(x)}}{2}\bigl(g^{\mu\alpha}(x)g^{\nu\beta}(x) + g^{\mu\beta}(x)g^{\nu\alpha}(x) - g^{\mu\nu}(x)g^{\alpha\beta}(x)\bigr)\tilde{\delta}a(x,x') \nonumber\\
	   G_{22}(x,x') &= \sqrt{g(x)}\tilde{\delta}(x,x') \nonumber \\
	   G_{33}(x,x') &= \sqrt{g(x)}F^{-1}\tilde{\delta}(x,x')
    \end{align}
    A constant background field implies we have a gauge fixing term given by (\hyperref[sgf]{4.10}). Using gauge condition we find:
    \begin{equation*}
        Q_{\alpha\beta} = -2\eta_{\alpha\beta}\Box + \kappa^2\partial_\alpha\bar\chi\partial_\beta\bar\chi + \kappa^2 F^{-1}[\bar\phi]\partial_\alpha\bar\phi\partial_\beta\bar\phi 
    \end{equation*}
    which in the constant background field makes no contribution.
    The connections arising out of field-space metric up to $\mathcal{O}(\kappa^2)$ are:
	\begin{align}
        (\Gamma^1_{11})^{ \mu\nu\rho\sigma}_{\lambda\tau}(x,x',x'') =& \tilde{\delta}(x'',x')\tilde{\delta}(x'',x)\kappa\left(-\delta^{(\mu}_{(\lambda}g^{\nu)(\rho}\delta^{\sigma)}_{\tau)}\right. +\dfrac{1}{4}g^{\mu\nu}\delta^{\rho}_{(\lambda}\delta^{\sigma}_{\tau)}+\dfrac{1}{4}g^{\rho\sigma}\delta^{\mu}_{(\lambda}\delta^{\nu}_{\tau)}\nonumber\\ &\qquad\qquad\qquad\qquad\qquad\qquad+\left.\dfrac{1}{4}g_{\lambda\tau}g^{\mu(\rho}g^{\sigma)\nu} -\dfrac{1}{8}g_{\lambda\tau}g^{\mu\nu}g^{\rho\sigma}\right) \nonumber\\
	    (\Gamma^1_{22})_{\mu\nu}(x,x',x'') =&\dfrac{1}{4}\tilde{\delta}(x'',x')\tilde{\delta}(x'',x)\kappa g_{\mu\nu} \nonumber\\
        (\Gamma^2_{12})^{\mu\nu}(x,x',x'')=&\dfrac{1}{4}\tilde{\delta}(x'',x')\tilde{\delta}(x'',x)\kappa g^{\mu\nu} \nonumber\\	        (\Gamma^1_{33})_{\mu\nu}(x,x',x'') =&\dfrac{1}{4}\tilde{\delta}(x'',x')\tilde{\delta}(x'',x)\kappa g_{\mu\nu}\left(1-\dfrac{1}{4\sqrt{6}}\kappa\chi\right)\nonumber\\	        (\Gamma^3_{13})^{\mu\nu}(x,x',x'')=&\dfrac{1}{4}\tilde{\delta}(x'',x')\tilde{\delta}(x'',x)\kappa g^{\mu\nu} \nonumber\\
	    \Gamma^2_{33}(x,x',x'') = &\tilde{\delta}(x'',x')\tilde{\delta}(x'',x)\left(\dfrac{\kappa}{2\sqrt{6}}-\dfrac{\kappa^2\chi}{12}\right)\nonumber\\
	    \Gamma^3_{22}(x,x',x'') = &-\dfrac{\kappa}{2\sqrt{6}}\tilde{\delta}(x'',x')\tilde{\delta}(x'',x)
	\end{align}
	Note that here we shall avoid using $\beta$ since this doesn't involve any comparison to some other existing literature. Following the procedure described previously in Section \hyperref[modela]{4.1}, we find the first term of (\hyperref[nonmineff]{3.11}) quadratic in background fields,
	\begin{multline}
            W(0) = \frac{1}{16\pi^2\epsilon}\int d^4x \Bigg(-\dfrac{1}{2}m_o^4 + \sqrt{\frac{3}{2}}\kappa m_o^4\chi - \dfrac{41}{48}\kappa^2 m_o^4 \chi^2 - \dfrac{1}{4}\kappa^2 m_o^4\xi\chi^2 - \dfrac{3}{4}\kappa^2 m_o^4\xi^2\chi^2 \\- m_o^2\zeta^2\phi^2- \dfrac{3}{4}\kappa^2 m_o^4\xi\phi^2 - \dfrac{3}{2}\kappa^2 m_o^4\xi^2\phi^2\Bigg) 
    \end{multline}
    Operator $\hat K(\nabla)$ takes the form,
    \begin{equation}
    \hat K(\nabla) = \begin{pmatrix}
    \delta^\alpha_{(\mu}\delta^\beta_{\nu)}\Box+2\alpha\delta^{(\alpha}_{(\mu}\nabla^{\beta)}\nabla_{\nu)}-\alpha\eta^{\alpha\beta}\nabla_{\mu}\nabla_{\nu} & 0 & 0 \\ 0 & \Box & 0 \\ 0 & 0 & \Box
    \end{pmatrix}
    \end{equation}
    Just as in the previous models, contribution from the second term in (\hyperref[nonmineff]{3.11}), upto quadratic order in background field, is, found to be: $-\dfrac{\alpha\kappa^2 m_o^4 \chi^2}{128\pi^2\epsilon}$ . In the limit $\alpha \rightarrow -1$, we get,
    \begin{multline}\label{gam1div3}
        \Gamma^{(1)}_{div} = \frac{-1}{16\pi^2\epsilon}\int d^4x \Bigg(\dfrac{1}{2}m_o^4 - \sqrt{\frac{3}{2}}\kappa m_o^4\chi + \dfrac{35}{48}\kappa^2 m_o^4 \chi^2 + \dfrac{1}{4}\kappa^2 m_o^4\xi\chi^2 + \dfrac{3}{4}\kappa^2 m_o^4\xi^2\chi^2 \\+ m_o^2\zeta^2\phi^2+ \dfrac{3}{4}\kappa^2 m_o^4\xi\phi^2 + \dfrac{3}{2}\kappa^2 m_o^4\xi^2\phi^2\Bigg)
    \end{multline}
    The results, here, were found to be in complete agreement with the perturbative approach, as outlined in Ref. \cite{mackay}, thus verifying that the extended Schwinger-DeWitt technique indeed performs as intended, and in turn validates the results obtained in Section \hyperref[modela]{4.1}.

    \section{Renormalization}\label{renorm}
    For the purpose of renormalization, we write the bare quantities occuring in action as,
    \begin{equation}
        m_{bare}^2 = m^2 + \delta m^2, \qquad
        \lambda_{bare} = l^\epsilon(\lambda + \delta\lambda)
    \end{equation}
    where $m$ and $\lambda$ represent, respectively, the renormalized mass and renormalized coupling parameters, and $\delta m$ and $\delta \lambda$ represent their corresponding counterterms. Now, the counterterm part of classical action for flat spacetime quartic in fields is given by,
    \begin{equation}
        \delta S = \int d^4x\Bigg(\dfrac{1}{2}\delta m^2\phi^2 + \dfrac{1}{24}\delta\lambda\phi^4\Bigg)
    \end{equation}
    From (\hyperref[gam1div1]{4.27}) and (\hyperref[gam1div2]{4.53}), it is seen that the divergent terms appearing in one loop part of effective action (excluding field independent terms) are of the form:
    \begin{equation}
        \Gamma^{(1)}_{div} = \int d^4x(A\phi^2 + B\phi^4)
    \end{equation}
     To absorb the divergent part of effective action, we require,
    \begin{equation}
        \delta m^2 = -2A,\qquad
        \delta \lambda = -24B 
    \end{equation}
    
    \subsection{Scalar Fields and Gravity}
    \subsubsection{Minimal Coupling}
    \noindent We can read off the coefficients $A$ and $B$ from (\hyperref[gam1div1]{4.27}):
    \begin{equation}
    A = \frac{1}{16\pi^2\epsilon}\left(\dfrac{5}{8}\kappa^2m^4 - \dfrac{1}{2}m^2\lambda\right),\qquad B = \frac{1}{16\pi^2\epsilon}\left(\dfrac{13}{48}m^2\kappa^2\lambda - \dfrac{1}{8}\lambda^2\right)
    \end{equation}
    The counterterms are, therefore,
    \begin{align}
        \delta m^2 = \frac{1}{16\pi^2\epsilon}\left(m^2\lambda-\dfrac{5}{4}\kappa^2m^4\right),\qquad
        \delta \lambda = -\frac{1}{16\pi^2\epsilon}\left(\dfrac{13}{2}m^2\kappa^2\lambda + 3\lambda^2\right)
    \end{align}
    \subsubsection{Non-Minimal Coupling}
    \noindent The coefficients read off from (\hyperref[gam1div2]{4.53}) are:
    \begin{align}
    A &= -\frac{1}{16\pi^2\epsilon}\left(\dfrac{1}{2}m^2\lambda + \dfrac{5}{8}\kappa^2m^4-2m^4\kappa^2\xi + 3m^4\kappa^2\xi^2\right)\nonumber\\
    B &= \frac{1}{16\pi^2\epsilon}\left(\dfrac{13}{48}m^2\kappa^2\lambda -\dfrac{1}{8}\lambda^2 - \dfrac{31}{24}m^2\kappa^2\lambda\xi + \dfrac{5}{2}m^2\kappa^2\lambda\xi^2\right)
    \end{align}
    And the corresponding counterterms are:
    \begin{align}
        \delta m^2 &= \frac{1}{16\pi^2\epsilon}\left(m^2\lambda - \dfrac{5}{4}\kappa^2m^4+4m^4\kappa^2\xi - 6m^4\kappa^2\xi^2\right)\nonumber\\
        \delta \lambda &= -\frac{1}{16\pi^2\epsilon}\left(\dfrac{13}{2}m^2\kappa^2\lambda +3\lambda^2 + 31m^2\kappa^2\lambda\xi - 60m^2\kappa^2\lambda\xi^2\right)
    \end{align}
    Note again that the calculations throughout this work have been carried out for a constant background scalar field. In this limit, the mass correction for the minimal coupling case is in agreement with the results found in Refs.  \cite{mackay,ashish,rodi}, and a similar agreement can be found with the results in Ref. \cite{stein} if instead of $\Gamma^{(1)}_{div}$, we read the coefficients $A$ and $B$ from $W(0)$.
    
    \subsection{Starobinsky Model with additional Scalar Field}
     Consider the action (\hyperref[actionb]{4.57}) expanded in powers of $\kappa$:
    
    \begin{multline}\label{expactionb}
        S = \int d^4x\sqrt{g(x)}\Bigg(\dfrac{1}{2}m_o^2\chi^2+\dfrac{1}{2}\chi^2\phi^2\zeta^2 + \dfrac{1}{4}\lambda\phi^4 - \dfrac{2R}{\kappa^2}+ \dfrac{1}{2}\partial_\mu\chi\partial^\mu\chi + \dfrac{1}{2}\partial_\mu\phi\partial^\mu\phi \\- \dfrac{\kappa}{2\sqrt{6}}\Big[m_o^2\chi^3 + 3m_o^2\xi\chi\phi^2 + \xi^2\chi^3\phi^2 + \lambda\chi\phi^4+ 3\zeta^2\xi\chi\phi^4+ \chi\partial_\mu\phi\partial^\mu\phi\Big] \\+ \dfrac{\kappa^2}{144}\Big[7m_o^2\chi^4 + 54m_o^2\xi\chi^2\phi^2 + 7\zeta^2\chi^4\phi^2 + 27m_o^2\xi^2\phi^4 \\+ 12\lambda\chi^2\phi^4 + 54\zeta^2\xi\chi^2\phi^4 + 27\zeta^2\xi^2\phi^6 + 6\chi^2\partial_\mu\phi\partial^\mu\phi\Big] + \mathcal{O}(\kappa^3) \Bigg)
    \end{multline}
    Note that there are no terms that are linear as well as dependent only on $\chi$ present in the expression above. However, from (\hyperref[gam1div3]{4.63}) we see that divergence comes from a term linear in $\chi$ and hence cannot be absorbed by counterterms present in the classical action. The theory is, therefore, non-renormalizable. The origin of the term linear in $\chi$ is the cubic self-interaction of the massive field $\chi$ appearing in the action above.
    
    \section{Verifications}\label{verify}
    
    \subsection{Non-Covariant Effective Action} \label{secncea}
     We now verify our results in Section \hyperref[modela]{4.1} Eqs. (\hyperref[w0]{4.22}, \hyperref[excconnec]{4.23}, and \hyperref[gam1div1]{4.27}) against \cite{stein} where the authors have carried out similar computations for a general Lagrangian coupling scalar fields with gravity in the Jordan frame. The authors claim to only work with non-covariant effective action obtained using the background field method for gauge theories, as discussed in Ref. \cite{diag}. Note again that by non-covariance in the context of this paper, we imply with respect to field reparametrizations. We wish to compare our results for such a calculation, and then list all the terms that would appear using VDW's gauge condition independent result. The action considered in Ref. \cite{stein} is,
    \begin{equation}\label{steinaction}
       S = \int d^4x \sqrt{g(x)}\Big[U(\varphi)R -\dfrac{1}{2}G(\varphi)g^{\mu\nu}\nabla_\mu\Phi^a\nabla_\nu\Phi_a - V(\varphi)\Big]
    \end{equation}
    for some functions $U(\varphi)$, $V(\varphi)$ and $G(\varphi)$, where $\varphi = \sqrt{\delta_{\mu\nu}\Phi^a\Phi^b}$, with $a,b=1,...,N$.\\
    Quoting its main result, while considering a flat spacetime background and assuming a constant scalar field  background, the one-loop divergence in the effective action turns out to be:
    \begin{equation}\label{steingam}
       \Gamma^{(1)}_{div} = \int d^4x \dfrac{\alpha_1}{32\pi^2(2-\omega)}
    \end{equation}
    where $\alpha_1$ is given by,
    \begin{multline}\label{alpha1}
        \alpha_1 = V^2\Bigg[\dfrac{2s^2(U')^4}{U^4} - \dfrac{2s(U')^2}{U^3} + \dfrac{5}{U^2}\Bigg]+VV'\Bigg[-\dfrac{8s^2(U')^3}{U^3} + \dfrac{4sU'}{U^2}\Bigg]+VV''\dfrac{2s^2(U')^2}{U^2}\\+(V')^2\Bigg[\dfrac{8s^2(U')^2}{U^2} -\dfrac{2s}{U} + \dfrac{N-1}{2G^2\phi^2}\Bigg] - V'V''\dfrac{4s^2U'}{U} + \dfrac{1}{2}(V'')^2s^2
    \end{multline}
    where $s = \dfrac{U}{GU+3(U')^2}$. Comparing (\hyperref[steinaction]{6.2}) with (\hyperref[minimalaction]{4.1}), we get for a single field $\phi$,
    \begin{equation}
        G(\phi) = 1,\qquad 
        U(\phi) = \dfrac{2}{\kappa^2},\qquad
        V(\phi) = -\dfrac{1}{2}m^2\phi^2 - \dfrac{1}{4!}\lambda\phi^4,\qquad
        s = 1
    \end{equation}
    Substituting these in (\hyperref[alpha1]{6.4}) and (\hyperref[steingam]{6.3}), we get,
    \begin{equation}
        \Gamma^{(1)}_{div} = \frac{-1}{16\pi^2\epsilon}\int d^4x \Bigg(\dfrac{1}{2}m^4 +\dfrac{1}{2}m^2\phi^2\lambda - \kappa^2m^4\phi^2-\dfrac{1}{3}m^2\kappa^2\lambda\phi^4 + \dfrac{1}{8}\lambda^2\phi^4\Bigg)\label{steinmainresult}
    \end{equation}
    which is the same as (\hyperref[excconnec]{4.23}). (Note: The difference in signs arises from the fact that action defined in (\hyperref[steinaction]{6.2}) carries an opposite sign compared to action defined in (\hyperref[minimalaction]{4.1})).
    
    For a non-minimal coupling case, comparing (\hyperref[steinaction]{6.2}) with (\hyperref[nonminaction]{4.28}), we get, upto $\mathcal O(\kappa^2)$,
    \begin{equation}
        U(\phi) = \dfrac{2}{\kappa^2} -\dfrac{1}{2}\xi\phi^2,\quad
        V(\phi) = -\dfrac{1}{2}m^2\phi^2 - \dfrac{1}{24}\lambda\phi^4,\quad
        G(\phi) = 1,\quad
        s = 1-\dfrac{3}{2}\xi^2\kappa^2\phi^2
    \end{equation}
    Substituting these in (\hyperref[alpha1]{6.4}) and (\hyperref[steingam]{6.3}), we get,
     \begin{multline}
        \Gamma^{(1)}_{div} = \frac{-1}{16\pi^2\epsilon}\int d^4x\Bigg(
        \dfrac{1}{2}m^4 - \kappa^2m^4\phi^2 +\dfrac{1}{2}m^2\phi^2\lambda+2m^4\kappa^2\xi\phi^2 - \dfrac{3}{2}m^4\kappa^2\xi^2\phi^2 - \dfrac{1}{3}m^2\lambda\kappa^2\phi^4\\ + \dfrac{1}{8}\lambda^2\phi^4 + \dfrac{4}{3}m^2\kappa^2\lambda\xi\phi^4 - \dfrac{3}{2}m^2\kappa^2\lambda\xi^2\phi^4\Bigg)
    \end{multline}
    which matches with (\hyperref[nonminbetazero]{4.51}).
    
    We, thus, conclude that the authors in Ref. \cite{stein} have indeed employed a non-covariant, background field approach, evident by the identification of our ordinary derivative results with theirs.  As such, the results in this paper can be seen as a covariant extension of the results presented in Ref. \cite{stein}.
    
     \subsection{On-Shell Effective Action} \label{secosea}
     While a non-trivial spacetime background is expected to yield a wider variety of corrections on-shell, due to constraints posed by computational complexity of assuming such a general case, we limit ourselves to a weak gravitational limit. Keeping in mind that the results in this limit are expected to be trivial, we present them anyway to verify the agreement of the covariant technique with the non-covariant results on-shell.
     
     Let us represent the models in Section \hyperref[modela]{4.1} in a general form as:
     \begin{equation}\label{6.1}
        S = \int d^4x \sqrt{g(x)}\Big[-U(\varphi)R +\dfrac{1}{2}G(\varphi)g^{\mu\nu}\nabla_\mu\Phi^a\nabla_\nu\Phi_a + V(\varphi)\Big]
     \end{equation}
     The classical equation of motion for metric field and scalar field are respectively (see Ref. \cite{steinJE}):
     \begin{equation}\label{6.2}
       R_{\alpha\beta}-\dfrac{1}{2}g_{\alpha\beta}R = \dfrac{G+2U''}{2U}\phi_{,\alpha}\phi_{,\beta}-\dfrac{G+4U''}{4U}g_{\alpha\beta}(\nabla\phi)^2+\dfrac{U'}{U}\phi_{;\alpha\beta}-\dfrac{U'}{U}g_{\alpha\beta}\Box\phi-\dfrac{1}{2}g_{\alpha\beta}\dfrac{V}{U}   
     \end{equation}
     \begin{equation}\label{6.3}
          \Box\phi = -\dfrac{U'}{G}R-\dfrac{1}{2}\dfrac{G'}{G}(\nabla\phi)^2+\dfrac{V'}{G}
     \end{equation}
     Under the simplifying conditions of flat spacetime background and constant background scalar field that we have employed throughout the paper these equations reduce to:
     \begin{equation}\label{6.4}
        V = 0,\qquad
        V'=0
     \end{equation}
     For the specific models in Section \hyperref[modela]{4.1}, these read:
     \begin{equation}\label{6.5}
          \dfrac{\phi^2}{2}\Big[m^2+\dfrac{1}{12}\lambda\phi^2\Big]=0,\qquad
         \phi\Big[m^2+\dfrac{1}{6}\lambda\phi^2\Big]=0
     \end{equation}
      whose only solution is:
      \begin{equation}\label{6.6}
          \phi = 0
      \end{equation}
      Thus, the on-shell one-loop effective action is (\hyperref[gam1div1]{4.27} and \hyperref[gam1div2]{4.53}):
      \begin{equation}\label{6.7}
      \Gamma^{(1)}_{div} = -\int d^4x\dfrac{m^4}{32\pi^2\epsilon}
      \end{equation}
      We have, thus, that both non-covariant (\hyperref[excconnec]{4.23} and \hyperref[nonminbetazero]{4.51}) and  covariant (\hyperref[gam1div1]{4.27} and \hyperref[gam1div2]{4.53}) effective actions lead to the same result on-shell. The result (\hyperref[6.7]{6.7}) is also in complete agreement with Ref. \cite{steinJE} when (\hyperref[6.4]{6.4}) along with solution (\hyperref[6.6]{6.6}) are used in Eq. (4) of Ref. \cite{steinJE}.

    \section{Discussion and Conclusions} \label{result}
    We derived the one-loop covariant effective action using the generalized Schwinger-DeWitt approach. In doing so, we get new terms corresponding to Vilkovisky-DeWitt connections (which were absent in non-covariant calculations of Ref. \cite{stein}) and the derived field-space metric (which are absent in perturbative calculations of Ref. \cite{mackay,gali} due to trivial choice of field-space metric). Moreover, in contrast to the perturbative approach, results obtained using heat kernel approach include the zero-point correction terms like $-\dfrac{m^4}{32\pi^2\epsilon}$ present in equations (\hyperref[gam1div1]{4.27}), (\hyperref[gam1div2]{4.53}), and  (\hyperref[gam1div3]{4.63}). Such corrections only cause a shift in vacuum energy, which does not contribute to the $\hat{S}$-matrix.
    
	Tallying the results of our calculations with the extensive literature on the subject, we first compared our output for the models considered in Section \hyperref[modela]{4.1} with the calculations carried out in Ref. \cite{mackay}, where the authors have employed the standard perturbative approach. Upon inspection, we found that for a minimal coupling, the results between the two methods matched exactly. However, in the presence of a non-minimal coupling term, the results differed.
	
	Comparing (\hyperref[gam1div2]{4.53}) with the output of Ref. \cite{mackay}, we found that in the limit $\xi \rightarrow 0$, the results are in agreement. However, for non-zero $\xi$, we find some additional terms namely,
	\begin{equation}
	  -\dfrac{m^4\kappa^2\xi\phi^2}{8\pi^2\epsilon} + \dfrac{3m^4\kappa^2\xi^2\phi^2}{16\pi^2\epsilon}\nonumber
	\end{equation}
	These terms are not in agreement with Ref. \cite{mackay} where the same type of terms enter with different coefficients. On investigating the issue, we discovered that the source of the disagreement was the choice of configuration-space metric used to carry out the relevant computations. The authors had chosen a diagonal field-space metric for interactive theory, with the Wheeler-DeWitt metric (\hyperref[gmetric]{4.2}) serving as the gravity part. This resulted in an incomplete picture since the correct metric for such a theory should instead be (\hyperref[nonminmetric]{4.42}). We believe this happened because the authors had chosen to separate the action into two separable parts involving gravity and scalar field, when the coupling term, present in the scalar part, must also contribute to the configuration-space metric in a non-trivial way, giving rise to off-diagonal terms.
	
	We further evaluated the non-covariant divergences in one-loop effective action in Section \hyperref[secncea]{6.1} and showed them to be in agreement with the previous calculations in Ref. \cite{stein}. We also presented an on-shell calculation of divergences in one-loop effective action in Section \hyperref[secosea]{6.2} and showed that the on-shell results match for covariant as well as non-covariant methods, which is in agreement with Ref. \cite{steinJE}. The consistency of the results obtained from the heat kernel technique with the results from \cite{mackay} and \cite{stein} in different limits also proves that the corrections presented here are not because of some byproduct of the technique itself, and appear as extensions of existing and established results.
    
    It must be noted that the local Schwinger-DeWitt expansion used throughout this paper is only applicable for cases where the fields are slow-varying and of small amplitude as compared to the mass-scale of the theory. Our choice of constant scalar field background is thus sound and fits in with the larger goal of extending the calculations to include time-dependent backgrounds and extracting relevant physical results from more cosmological models.
    
    An ideal application of this method would be in the context of gravitational corrections in the FLRW background which could lead to interesting studies of quantum gravitational effects in the inflationary regime, and is part of our future plans. In Section \hyperref[modelb]{4.2}, we performed the calculation for an action involving two scalar fields. This can, in principle, be extended to include a multiplet of fields involving scalar, tensor, as well as fermionic fields (Refer to Ref. \cite{karam2} for more information on the extension of the formalism to fermionic theories). Thus, the computational framework developed in this paper can be applied to far more complicated models which involve non-linearity, like in model \hyperref[modelb]{4.2}, or non-trivial field-space metric, like the models in Section \hyperref[modela]{4.1}, or both. This flexibility and range of the generalized Schwinger-DeWitt technique could help us explore a wide variety of cosmological models as part of our future work.
    
    \section*{Acknowledgement}
	The authors of this paper would like to thank D. J. Toms for his invaluable insight. The calculations in this paper have been carried out in MATHEMATICA using the xAct packages xTensor \cite{xtensor} and xPert \cite{xpert}. This work is partially supported by DST (Govt. of India) Grant No. SERB/PHY/2017041.

	\appendix
	\section{Heat Kernel Coefficients}\label{coeff}
	For an operator of the form (\hyperref[min]{3.5}), the heat kernel coefficients relevant for calculations in four dimensions are listed below:
    \begin{align}
        \hat a_0(x,x)=& \hat 1,   \\
        \hat a_1(x,x)=& \frac{1}{6}R\hat 1-\hat P, \\
        \hat a_2(x,x)=&\Big\{-\frac{1}{30}\Box R+\frac{1}{72}R^2+\frac{1}{180}(R_{\alpha\beta\gamma\delta}R^{\alpha\beta\gamma\delta}-R_{\mu\nu} R^{\mu\nu})\Big\}\hat 1 \nonumber\\&+\frac1{12}\hat{\cal R}_{\mu\nu}\hat{\cal R}^{\mu\nu} +\frac{1}{2}\hat{P}^2-\frac{1}{6}R\hat P+\frac{1}{6}\Box \hat{P}
    \end{align}
    Here, the coefficients are presented in their respective coincidence limits $(x'\to x)$. A functional trace (Tr) over the heat kernel $\hat K(s|x,x')$ would mean evaluating the trace of the field-space operator $\hat K$ followed by a spacetime integration in its coincidence limit.
    \begin{equation}
        \text{Tr}{\hat{K}(s|x,x')} = \int dv_x \;\text{tr}{\hat{K}(s|x,x)}
    \end{equation}

   \section{Killing's Equations}\label{Killing}
    An important property of the field-space metric that allows us to apply the analysis listed in this paper is that, combined with the gauge transformation acting as a Killing's vector, it must satisfy Killing's equation. We will show that the metrics chosen in models \hyperref[modela1]{4.1.1}, \hyperref[modela2]{4.1.2} and \hyperref[modelb]{4.2} indeed do so. Let us choose the following form of metric components:
\begin{align}
    (G_{11}^{\mu\nu\rho\sigma})(y,z) &= F[\phi(y)]\sqrt{g(y)}\left(g^{\mu(\rho}(y)g^{\sigma)\nu}(y) -\dfrac{1}{2}g^{\mu\nu}(y)g^{\rho\sigma}(y)\right)\tilde\delta(y,z)\nonumber\\
        (G_{12})^{\rho\sigma}(y,z) &= \sqrt{g(y)}H[\phi(y)]g^{\rho\sigma}(y)\tilde\delta(y,z)\nonumber\\
        G_{22}(y,z) &= \sqrt{g(y)}J[\phi(y)]\tilde\delta(y,z)\label{eq:B.1}
    \end{align}
    The gauge transformations are generated by,
    \begin{align}
        K^1_{\lambda\tau\alpha}(x,x') &= -(\partial_\alpha g_{\lambda\tau}(x) + g_{\alpha\lambda}(x)\partial_\tau + g_{\alpha\tau}(x)\partial_\lambda)\tilde\delta(x,x')\nonumber\\
        K^2_{\alpha}(x,x') &= -\partial_\alpha\phi(x)\tilde\delta(x,x')\label{eq:B.2}
    \end{align}
    Killing's equation is given by,
    \begin{equation}
        \label{eq:B.3}
    K^k_\alpha g_{mn,k} + K^k_{\alpha,n}g_{km} + K^k_{\alpha,m}g_{kn} = K^i_\epsilon g_{im}B^\epsilon_{\ \alpha n} + K^i_\epsilon g_{in}B^\epsilon_{\ \alpha m}
        \end{equation}
    To facilitate the evaluation of the terms in the equation above, we define a field independent test function $R^\alpha(x)$. By the end we will show that:
    \begin{equation}
        \label{eq:B.4}
   R^\alpha(K^k_\alpha g_{mn,k} + K^k_{\alpha,n}g_{km} + K^k_{\alpha,m}g_{kn})= 0
        \end{equation}
        i.e. Killing's equation is satisfied for $B^\epsilon_{\ \alpha m} = 0$. Consider the term  $R^\alpha K^k_\alpha$ for $k = \{g_{\lambda\tau}(x)\}$:
        \begin{equation*}
           R^\alpha K^k_\alpha = \int d^dx'K^1_{\lambda\tau\alpha}(x,x')R^\alpha(x')
           =-\int d^dx'((g_{\lambda\tau,\alpha}(x) + g_{\alpha\lambda}(x)\partial_\tau + g_{\alpha\tau}\partial_\lambda(x))\tilde\delta(x,x'))R^\alpha(x')
        \end{equation*}
    On integrating by parts we have,
    \begin{equation}
        \label{eq:B.5}
         R^\alpha K^k_\alpha = -\nabla^x_\lambda R_\tau(x)
 - \nabla^x_\tau R_\lambda(x)   
    \end{equation}
    where a superscript $x$ has been attached indicating the spacetime point with respect to which covariant derivative is to be taken.\\
    For $k = \{\phi(x)\}$, we have:
    \begin{equation}
        R^\alpha K^k_\alpha = \int d^dx'K^2_{\alpha}(x,x')R^\alpha(x')
        =-\int d^dx' \partial_\alpha\phi(x)\tilde\delta(x,x')R^\alpha(x')
        =-\partial_\alpha\phi(x)R^\alpha(x)\label{eq:B.6}
    \end{equation}
    \textbf{Metric component: } $(G_{11}^{\mu\nu\rho\sigma})(y,z)$\\\\
    For the metric component  $(G_{11}^{\mu\nu\rho\sigma})(y,z)$, we associate $m = \{g_{\mu\nu}(y)\}$ and $n=\{g_{\rho\sigma}(z)\}$. Then, for $k = \{g_{\lambda\tau}(x)\}$ we have:
    \begin{align*}
        g_{mn,k} =& \dfrac{\delta (G_{11})^{\mu\nu\rho\sigma}(y,z)}{\delta g_{\lambda\tau}(x)}\\ =& \Big\{-\dfrac{1}{2}(g^{\mu(\lambda}g^{\tau)\rho}g^{\nu\sigma} + g^{\mu\rho}g^{\nu(\lambda}g^{\tau)\sigma} + g^{\mu(\lambda}g^{\tau)\sigma}g^{\nu\rho} + g^{\mu\sigma}g^{\nu(\lambda}g^{\tau)\rho} - g^{\mu(\lambda}g^{\tau)\nu}g^{\rho\sigma} \\&- g^{\mu\nu}g^{\rho(\lambda}g^{\tau)\sigma}) + \dfrac{1}{2}g^{\lambda\tau}(y)\Big(g^{\mu(\rho}g^{\sigma)\nu} - \dfrac{1}{2}g^{\mu\nu}g^{\rho\sigma}\Big)\Big\}_yF[\phi(y)]\sqrt{g(y)}\tilde\delta(y,x)\tilde\delta(y,z)
    \end{align*}
  where we have again attached a subscript $y$ to indicate the spacetime argument of expressions enclosed within the brackets. Using the expression above with (\ref{eq:B.5}), we get for $k = \{g_{\lambda\tau}(x)\}$,
  \begin{align}
      R^\alpha K^k_\alpha g_{mn,k} =& \sqrt{g(y)}\tilde\delta(y,z)F[\phi(y)]\Big\{-g^{\mu(\rho}g^{\sigma)\nu}\nabla_\alpha R^\alpha+\dfrac{1}{2}g^{\mu\nu}g^{\rho\sigma}\nabla_\alpha R^\alpha + g^{\mu\rho}\nabla^{(\nu} R^{\sigma)}\nonumber \\&+ g^{\mu\sigma}\nabla^{(\nu} R^{\rho)} + g^{\nu\rho}\nabla^{(\mu}R^{\sigma)} + g^{\nu\sigma}\nabla^{(\mu}R^{\rho)} - g^{\mu\nu}\nabla^{(\sigma}R^{\rho)} - g^{\rho\sigma}\nabla^{(\mu}R^{\nu)}\Big\}_y      \label{eq:B.7}
\end{align}
For $k = \{\phi(x)\}$ using (\ref{eq:B.6}), we get,
\begin{align}
     R^\alpha K^k_\alpha g_{mn,k} &=   -\int d^dx \partial_\alpha\phi(x)R^\alpha(x)\dfrac{\delta (G_{11}^{\mu\nu\rho\sigma})(y,z)}{\delta \phi(x)}\nonumber \\&= -\sqrt{g(y)}\partial_\alpha\phi(y)R^\alpha(y)\Big\{F'[\phi]\Big(g^{\mu(\rho}g^{\sigma)\nu} -\dfrac{ g^{\mu\nu}g^{\rho\sigma}}{2}\Big)\Big\}_y\tilde\delta(y,z)\label{eq:B.8}
    \end{align}
    Using (\ref{eq:B.7}) and (\ref{eq:B.8}) we get:
 \begin{multline}
      R^\alpha K^k_\alpha g_{mn,k} = \sqrt{g(y)}\tilde\delta(y,z)\Big[F[\phi(y)]\Big\{-g^{\mu(\rho}g^{\sigma)\nu}\nabla_\alpha R^\alpha+\dfrac{1}{2}g^{\mu\nu}g^{\rho\sigma}\nabla_\alpha R^\alpha + g^{\mu\rho}\nabla^{(\nu} R^{\sigma)} \\+ g^{\mu\sigma}\nabla^{(\nu} R^{\rho)} + g^{\nu\rho}\nabla^{(\mu}R^{\sigma)} + g^{\nu\sigma}\nabla^{(\mu}R^{\rho)} - g^{\mu\nu}\nabla^{(\sigma}R^{\rho)} - g^{\rho\sigma}\nabla^{(\mu}R^{\nu)}\Big\}_y\\ - \Big\{(\partial_\alpha\phi) R^\alpha F'[\phi]\Big(g^{\mu(\rho}g^{\sigma)\nu} -\dfrac{ g^{\mu\nu}g^{\rho\sigma}}{2}\Big)\Big\}_y\Big]     \label{eq:B.9}
\end{multline}
Now, consider the following term for $k = \{g_{\lambda\tau}(x)\}$:
\begin{equation}\label{eq:B.10}
    R^\alpha K^k_{\alpha,n} = (R^\alpha K^k_{\alpha})_{,n} = \dfrac{\delta(-\nabla_{(\lambda}R_{\tau)}(x))}{\delta g_{\rho\sigma}(z)}  
\end{equation}
To evaluate this we expand $\nabla_{(\lambda}R_{\tau)}$,
\begin{equation}
    \label{eq:B.11}
    \nabla_{(\lambda}R_{\tau)} = -2g_{\alpha(\tau}\partial_{\lambda)}R^\alpha - g_{\lambda\tau,\alpha}R^\alpha
\end{equation}
Using (\ref{eq:B.11}) in (\ref{eq:B.10}), we get:
\begin{equation}
    \label{eq:B.12}
     R^\alpha K^k_{\alpha,n} = \Big[-\delta^\rho_{(\alpha}\delta^\sigma_{\tau)}\partial_{\lambda}R^\alpha - \delta^\rho_{(\alpha}\delta^{\sigma}_{\lambda)}\partial_\tau R^\alpha - R^\alpha\delta^\rho_{(\lambda}\delta^\sigma_{\tau)}\partial_\alpha\Big]_x\tilde\delta(x,z)
\end{equation}
We can modify this as:
\begin{equation}\label{eq:B.13}
         R^\alpha K^k_{\alpha,n} = \Big[\Big(-\delta^\rho_{(\alpha}\delta^\sigma_{\tau)}\partial_{\lambda}R^\alpha - \delta^\rho_{(\alpha}\delta^{\sigma}_{\lambda)}\partial_\tau R^\alpha -2\delta^{(\rho}_{(\tau}\Gamma^{\sigma)}_{\ \lambda)\alpha}R^\alpha\Big) - \Big(R^\alpha\delta^\rho_{(\lambda}\delta^\sigma_{\tau)}\partial_\alpha  -2\delta^{(\rho}_{(\tau}\Gamma^{\sigma)}_{\ \lambda)\alpha}R^\alpha \Big)\Big]_x\tilde\delta(x,z)
\end{equation}
We can now use the identity:
\begin{equation}
    \label{eq:B.14}
    \Big[R^\alpha\delta^\rho_{(\lambda}\delta^\sigma_{\tau)}\nabla_\alpha\Big]_x\tilde\delta(x,x') = \Big[R^\alpha\delta^\rho_{(\lambda}\delta^\sigma_{\tau)}\partial_\alpha -2\delta^{(\rho}_{(\tau}\Gamma^{\sigma)}_{\ \lambda)\alpha}R^\alpha
    \Big]_x\tilde\delta(x,x')
\end{equation}
to get:
\begin{equation}
    \label{eq:B.15}
      R^\alpha K^k_{\alpha,n} = \Big[-\delta^\sigma_{(\tau}\nabla_{\lambda)}R^\rho -\delta^\rho_{(\tau}\nabla_{\lambda)}R^\sigma - R^\alpha\delta^\rho_{(\lambda}\delta^\sigma_{\tau)}\nabla_\alpha\Big]_x\tilde\delta(x,z)
\end{equation}
(Note: The identity (\ref{eq:B.14}) can be proven by multiplying an arbitrary symmetric test function $f_{\rho\sigma}(x')$ and integrating over $x'$ on both sides and showing they give the same result.)
For $k = \{\phi(x)\}$ we have:
\begin{equation}
    \label{eq:B.16}
     R^\alpha K^k_{\alpha,n} = (R^\alpha K^k_{\alpha})_{,n} = \dfrac{\delta(-(\partial_\alpha\phi(x))R^\alpha(x))}{\delta g_{\rho\sigma}(z)} = 0
\end{equation}
Combining (\ref{eq:B.15}) and (\ref{eq:B.16}), we get the second term of (\ref{eq:B.4}):
\begin{multline}
    R^\alpha K^k_{\alpha,n}g_{km} = \sqrt{g(y)}F[\phi(y)]\Big\{-g^{\nu(\rho}\nabla^\mu R^{\sigma)} - g^{\mu(\rho}\nabla^\nu R^{\sigma)} + g^{\mu\nu}\nabla^{(\sigma}R^{\rho)} \\- g^{\rho(\mu}g^{\nu)\sigma}R^\alpha\nabla_\alpha + \dfrac{1}{2}g^{\rho\sigma}g^{\mu\nu}R^\alpha\nabla_\alpha\Big\}_y\tilde\delta(y,z)    \label{eq:B.17}
\end{multline}
Similar calculations show:
\begin{multline}
    R^\alpha K^k_{\alpha,m}g_{kn} = \sqrt{g(z)}F[\phi(z)]\Big\{-g^{\sigma(\mu}\nabla^\rho R^{\nu)} - g^{\rho(\mu}\nabla^\sigma R^{\nu)} + g^{\rho\sigma}\nabla^{(\nu}R^{\mu)} \\- g^{\mu(\rho}g^{\sigma)\nu}R^\alpha\nabla_\alpha + \dfrac{1}{2}g^{\mu\nu}g^{\rho\sigma}R^\alpha\nabla_\alpha\Big\}_z\tilde\delta(z,y)    \label{eq:B.18}
\end{multline}
Using the identities:
\begin{align}
    f(x')\sqrt{g(x')}\delta(x,x') &= f(x)\sqrt{g(x)}\delta(x',x)\nonumber\\
    \Big[f^\alpha(x')\sqrt{g(x')}\nabla'_{\alpha}\Big]\tilde\delta(x',x) &= -\sqrt{g(x)}\Big[\nabla_\alpha f^\alpha(x) + f^\alpha(x)\nabla_\alpha\Big]\tilde\delta(x,x')\label{eq:B.19}
\end{align}
it is straightforward to see from (\ref{eq:B.9}), (\ref{eq:B.17}), and (\ref{eq:B.18}) that:
    \begin{equation*}
   R^\alpha(K^k_\alpha g_{mn,k} + K^k_{\alpha,n}g_{km} + K^k_{\alpha,m}g_{kn})= 0
        \end{equation*}
        
\noindent\textbf{Metric Component: }$(G_{12})^{\rho\sigma}(y,z)$\\\\
Here, we associate $m = \{\phi(y)\}$ and $n = \{g_{\rho\sigma}(z)\}$. For $k = \{g_{\lambda\tau}(x)\}$:
\begin{align}
    R^\alpha K^k_\alpha g_{mn,k} &=\int d^dx (-\nabla^x_{(\lambda}R_{\tau)}(x))\dfrac{\delta ((G_{12})^{\rho\sigma}(y,z))}{\delta g_{\lambda\tau}(x)} \nonumber\\&= \int d^dx (-\nabla^x_{(\lambda}R_{\tau)}(x))\sqrt{g(y)}H[\phi(y)]\Big\{-g^{\rho(\lambda}g^{\tau)\sigma} + \dfrac{1}{2}g^{\lambda\tau}g^{\rho\sigma}\Big\}_y\tilde\delta(y,x)\tilde\delta(y,z)\nonumber\\
    &=\sqrt{g(y)}\tilde\delta(y,z)H[\phi(y)]\Big\{\nabla^{\rho}R^\sigma + \nabla^\sigma R^\rho - \nabla_\alpha R^\alpha\Big\}_y\label{eq:B.20}
\end{align}
For $k = \{\phi(x)\}$,
\begin{align}
 R^\alpha K^k_\alpha g_{mn,k}  =& \int d^dx (-\partial_\alpha\phi(x)R^\alpha(x)))\dfrac{\delta ((G_{12})^{\rho\sigma}(y,z))}{\delta \phi(x)} \\=& -\sqrt{g(y)}\partial_\alpha\phi(y)R^\alpha(y)H'[\phi(y)]g^{\rho\sigma}(y)\tilde\delta(y,z)\label{eq:B.21}
 \end{align}
 Combining (\ref{eq:B.20}) and (\ref{eq:B.21}),
 \begin{align}
      R^\alpha K^k_\alpha g_{mn,k}  &= \sqrt{g(y)}\tilde\delta(y,z)\Big[H[\phi]\Big(\nabla^{\rho}R^\sigma + \nabla^\sigma R^\rho - \nabla_\alpha R^\alpha\Big) - \partial_\alpha\phi R^\alpha H'[\phi] g^{\rho\sigma}\Big]_y\label{eq:B.22}
 \end{align}
 For the second term of (\ref{eq:B.4}) only $k = \{g_{\lambda\tau}(x)\}$ can be taken. Using (\ref{eq:B.15}),
 \begin{align}
      R^\alpha K^k_{\alpha,n}g_{km} =& \int d^dx \sqrt{g(x)}H[\phi(x)]g^{\lambda\tau}(x)\tilde\delta(x,y)\Big\{-\delta^\sigma_{(\tau}\nabla_{\lambda)}R^\rho \\&\quad\quad\quad\quad\quad\quad\quad\quad\quad\quad-\delta^\rho_{(\tau}\nabla_{\lambda)}R^\sigma - R^\alpha\delta^\rho_{(\lambda}\delta^\sigma_{\tau)}\nabla_\alpha\Big\}_x\tilde\delta(x,z)\nonumber\\
      =&\sqrt{g(y)}H[\phi(y)]\Big\{-\nabla^\sigma R^\rho - \nabla^\rho R^\sigma - g^{\rho\sigma}R^\alpha\nabla_\alpha\Big\}_y\tilde\delta(y,z)
      \label{eq:B.23}
 \end{align}
For the third term of (\ref{eq:B.4}), only $k = \{\phi(x)\}$ can be taken,
\begin{align}
  R^\alpha K^k_{\alpha,m}g_{kn} &= \int d^dx\dfrac{\delta(-\nabla_\alpha\phi(x)R^\alpha(x))}{\delta \phi(y)}H[\phi(x)]g^{\rho\sigma}(x)\sqrt{g(x)}\tilde\delta(x,z)\nonumber\\
  &= -R^\alpha(z)H[\phi(z)]g^{\rho\sigma}(z)\sqrt{g(z)}\nabla^z_\alpha\{\tilde\delta(z,y)\}\label{eq:B.24}
  \end{align}
Using the identities (\ref{eq:B.19}) in (\ref{eq:B.23}), we get:
\begin{align}
     R^\alpha K^k_{\alpha,n}g_{km} =& \sqrt{g(y)}H[\phi(y)]\Big\{-\nabla^\sigma R^\rho - \nabla^\rho R^\sigma\Big\}_y\tilde\delta(y,z) \nonumber\\&\quad\quad+\sqrt{g(z)}g^{\rho\sigma}(z)H[\phi(z)]R^\alpha(z)\nabla^z_\alpha\{\tilde\delta(z,y)\}
      \nonumber\\&\quad\quad\quad\quad\quad\quad+\sqrt{g(z)}\tilde\delta(z,y)g^{\rho\sigma}(z)\nabla^z_\alpha\{H[\phi(z)]R^\alpha(z)\}\nonumber\\
     =& \sqrt{g(y)}H[\phi(y)]\Big\{-\nabla^\sigma R^\rho - \nabla^\rho R^\sigma + \nabla_\alpha R^\alpha\Big\}_y\tilde\delta(y,z) \nonumber\\
      &\quad\quad+\sqrt{g(z)}g^{\rho\sigma}(z)H[\phi(z)]R^\alpha(z)\nabla^z_\alpha\{\tilde\delta(z,y)\} \nonumber\\&\quad\quad\quad\quad\quad\quad+\sqrt{g(y)}\tilde\delta(y,z)g^{\rho\sigma}(y)R^\alpha(y)H'[\phi(y)]\partial_\alpha\phi(y)\label{eq:B.25}
\end{align}
On combining (\ref{eq:B.22}), (\ref{eq:B.24}), and (\ref{eq:B.25}), it is straightforward to see that (\ref{eq:B.4}) is satisfied.\\

\noindent\textbf{Metric Component: }$G_{22}(y,z)$\\\\
For this, we associate $m = \{\phi(y)\}$ and $n = \{\phi(z)\}$.
The first term in (\ref{eq:B.4}) for $k = \{g_{\lambda\tau}(x)\}$ is:
\begin{align}
    R^\alpha K^k_{\alpha}g_{mn,k}=& \int d^dx (-\nabla_{(\lambda}^xR_{\tau)}(x))\sqrt{g(y)}J[\phi(y)]g^{\lambda\tau}(y)\tilde\delta(y,z)\tilde\delta(y,x)\nonumber\\=& -\nabla^y_\alpha R^\alpha(y)\sqrt{g(y)}J[\phi(y)]\tilde\delta(y,z)\label{eq:B.26}
\end{align}
For $k = \{\phi(x)\}$,
\begin{align}
        R^\alpha K^k_{\alpha}g_{mn,k} &= \int d^dx (-\partial_\alpha\phi(x)R^\alpha(x))\sqrt{g(y)}J'[\phi(y)]\tilde\delta(y,z)\tilde\delta(y,x)\nonumber\\&= -\partial_\alpha\phi(y)R^\alpha(y)\sqrt{g(y)}J'[\phi(y)]\tilde\delta(y,z)\label{eq:B.27}
\end{align}
Combining (\ref{eq:B.26}) and (\ref{eq:B.27}),
\begin{equation}\label{eq:B.28}
    R^\alpha K^k_\alpha g_{mn,k} = -\sqrt{g(y)}\tilde\delta(y,z)\nabla^y_\alpha \{R^\alpha(y)J[\phi(y)]\} 
\end{equation}
For the second term in (\ref{eq:B.4}), only $k = \{\phi(x)\}$ can be taken.
\begin{align}
    R^\alpha K^k_{\alpha,n}g_{km} &= \int d^dx (-R^\alpha(x)\nabla^x_\alpha\{\tilde\delta(x,z)\})\sqrt{g(x)}J[\phi(x)]\tilde\delta(x,y)\nonumber\\&=-R^\alpha(y)J[\phi(y)]\sqrt{g(y)}\nabla^y_\alpha\{\tilde\delta(y,z)\}\label{eq:B.29}
\end{align}
For the third term also, only $k = \{\phi(x)\}$ can be taken.
\begin{equation}
     R^\alpha K^k_{\alpha,m}g_{kn} = -R^\alpha(z)J[\phi(z)]\sqrt{g(z)}\nabla^z_\alpha\{\tilde\delta(z,y)\}\label{eq:B.30}
\end{equation}
On using identities (\ref{eq:B.19}) in the equation above and then combining it with (\ref{eq:B.28}) and (\ref{eq:B.29}), it is seen that Killing's equation (\ref{eq:B.4}) is satisfied. It is therefore verified that all the components for the choice of metric (\ref{eq:B.1}) satisfy Killing's equation. Similar verification can be carried out for the field-space metric in model \hyperref[modelb]{4.2}.


\end{document}